\documentclass[twocolumn]{autart}    
\usepackage{amsfonts}
\usepackage{textcomp}
\usepackage{enumerate}
\usepackage{cite}
\usepackage{color}
\usepackage{amssymb}
\usepackage{graphicx}
\usepackage{algorithm}
\usepackage{algpseudocode}
\usepackage{bbm}
\usepackage{graphics}
\usepackage{epsfig}
\usepackage{multicol} 
\usepackage{stfloats}
\usepackage{appendix}
\usepackage{amsmath}
\usepackage[table]{xcolor}
\definecolor{revision}{RGB}{0,0,0} 
\usepackage{subcaption}
\usepackage{placeins}
\usepackage{url}
\usepackage{booktabs}
\usepackage{makecell}
\usepackage{multirow}
\usepackage{wrapfig}

\newtheorem{theorem}{Theorem}
\newtheorem{proposition}{Proposition}
\newtheorem{lemma}{Lemma}

\newtheorem{remark}{Remark}

\newtheorem{assumption}{Assumption}
\makeatletter

\newcommand{\Rmnum}[1]{\expandafter\@slowromancap\romannumeral #1@}
\makeatother                                   

\begin{document}

\begin{frontmatter}

\title{Secure State Estimation of Cyber-Physical Systems via  Gaussian Bernoulli Mixture Model} 

\author[address_a]{Xingzhou~Chen},~
\author[address_a]{Nachuan~Yang},~
\author[address_c]{Peihu~Duan},~
\author[address_d]{Shilei~Li}$^\dagger$,~
\author[address_a,address_e]{Ling~Shi}

\address[address_a]{Department of Electronic and Computer Engineering, The Hong Kong University of Science and Technology,\\
Clear Water
Bay, Kowloon, Hong Kong, China}

\address[address_c]{State Key Laboratory of CNS/ATM, Beijing Institute of Technology, Beijing 100081, China}

\address[address_d]{School of Automation, Beijing Institute of Technology, Beijing 100081, China}

\address[address_e]{Department of Chemical and Biological Engineering, The Hong Kong University of Science and Technology,\\Clear Water Bay, Kowloon, Hong Kong, China}


\begin{abstract}
The implementation of cyber-physical systems in real-world applications is challenged by safety requirements in the presence of sensor threats. Most cyber-physical systems, especially multi-sensor systems, struggle to detect sensor attacks when the attack model is unknown. In this paper, we tackle this issue by proposing a Gaussian-Bernoulli Secure (GBS) estimator, which transforms the detection problem into an optimal estimation problem concerning the system state and observation indicators. We solve the GBS estimator design problem from two complementary perspectives. On the one hand, the problem can be viewed as an observation-subset selection problem: sequential state estimation under different partial observations. On the other hand, the problem can be viewed as an observation-set update problem. {\color{revision}When a few indicators change in arbitrary order, a warm-started proximal gradient iteration updates the estimate efficiently, without re-solving the problem or enumerating subsets. We further establish a worst-case bound on the number of iterations for measurement selection. Moreover, we prove that, for arbitrary attack sequences, the mean-square estimation error admits a horizon-independent bound, provided that the retained clean measurements satisfy a window observability condition.} We evaluate the proposed GBS estimator from three perspectives: computational efficiency, detection performance, and estimation error.
\end{abstract}

\end{frontmatter}

\section{Introduction}
Cyber-Physical Systems (CPSs) are integrative frameworks that combine computing and communication capabilities to monitor or control entities within the physical world~\cite{giraldo2018survey,jensen2011model,zhou2024cybersecurity}. Over the past two decades, significant advancements in communication and computing technologies have catalyzed the wide applications for CPSs~\cite{chen2017applications}. The advancement in wireless communication technologies has enabled CPSs to transmit real-time signals across expansive physical networks~\cite{yu2016smart}. Furthermore, armed with advanced computational resources, CPS can optimize control strategies for complex systems online in real-time~\cite{zhang2022advancements}. This technological evolution broadens the scope of CPSs, impacting various sectors such as robot manipulators~\cite{ren2025high}, smart grids~\cite{he2016cyber}, and autonomous vehicles~\cite{kim2021cybersecurity}, to name just a few.

However, alongside the opportunities, CPSs face significant challenges, with security issues emerging as a paramount concern. A notable incident underscoring this concern involves the RQ-170, an American drone~\cite{ruegamer2015jamming}\cite{hartmann2013vulnerability}, which lost control in Iran due to a security breach. This security breach was executed through the manipulation of the drone's GPS signal, causing the UAV to land in a strategically targeted area. The reliance of many CPSs on remote signal transmission over vulnerable networks opens them up to sensory attacks. These attacks, through the injection of malicious data~\cite{mo2010false} or disruption of information transmission~\cite{ren2024optimal}, can significantly degrade a system's performance, e.g., maximizing the estimation error covariance~\cite{guo2016optimal}, amplifying the expected estimation error~\cite{zhang2015optimal1}. 

Model-based methods for detecting sensor attacks include $\chi^2$ detectors that monitor the Kalman innovation sequence~\cite{mo2013detecting}.
Interval anomaly detectors have also been studied in remote state estimation under deception attacks~\cite{zhou2023optimal}.
For attacks on actuators, Yang et al.~\cite{yang2023lasso} used LASSO to detect and identify compromised inputs.
Data-driven approaches include the graphical model of Lin et al.~\cite{lin2018tabor}, which was used to localize anomalies in a water treatment system.
Goh et al.~\cite{goh2017anomaly} used an LSTM recurrent neural network to learn temporal patterns in CPS measurements and identify sensor anomalies.

Apart from detecting attacks, designing resilient state estimators plays a critical role in mitigating sensor attacks. Mo et al.~\cite{mo2014secure} studied the case of multiple measurements subject to sensor attacks and designed an optimal estimator by solving a minimax optimization problem. For secure estimation in noiseless systems, Chang et al.~\cite{chang2018secure} formulated the problem as a classical error-correction problem and provided sufficient conditions for recovering the true states. {Chen et al.~\cite{chen2017maximum} proposed the maximum correntropy Kalman filter (MCKF), which replaces the MMSE criterion with the maximum correntropy criterion so that large-residual measurements are suppressed, leading to improved robustness against heavy-tailed noise and outliers.}

{\color{revision}\label{sec:related-comparison}
Existing approaches to secure state estimation and outlier detection exploit sparsity or reduce the influence of corrupted measurements. Sensor-subset selection methods identify a consistent set of measurements~\cite{shoukry2017,alessandri2016outliers}, but the number of candidate subsets grows combinatorially. Sparse-offset estimation~\cite{farahmand2011} relaxes the problem with an $\ell_1$ penalty, which biases the estimated corruption magnitudes and lets part of each large corruption leak into the state estimate. Binary indicators have also been incorporated into recursive estimation, as in the beta--Bernoulli filter of Wang et al.~\cite{wang2018robust}, but such a filter does not revisit earlier decisions when later measurements arrive. Building on these ideas, we associate each measurement with a binary indicator and jointly estimate the indicators and the state trajectory. Our approach aims to accelerate combinatorial subset search, mitigate the bias-induced accuracy loss of convex relaxation under structured attacks, and allow later measurements to inform earlier rejection decisions.}

{\color{revision}To address these problems, we propose the Gaussian--Bernoulli secure (GBS) estimator. The main contributions are as follows.
\begin{enumerate}
    \item We introduce a Gaussian--Bernoulli mixture model to capture discrete anomaly events in a system with continuous-valued states, without requiring prior knowledge of the attack model.
    \item We derive a closed-form Gaussian trajectory estimate for each fixed observation subset and accelerate updates under arbitrarily ordered subset changes via a composite convex formulation and proximal gradient descent.
    \item We bound the number of accepted indicator updates and establish a horizon-independent mean-square error bound when the retained clean measurements satisfy a window observability condition.
    \item We compare GBS with the $\chi^2$ detector, the CUSUM detector, the resilient estimator, and the sparse-offset estimator under persistent biases, gradual drifts, and sensor replay attacks. Without any attacker information, GBS generally achieves higher detection accuracy and lower estimation error.
\end{enumerate}}

The remainder of the paper is organized as follows. In Section~2, we introduce linear estimation as an application of Bayesian theory in Hidden Markov Chains. In Section~3, we describe the model assumptions and the problem we aim to solve. In Sections~4 and~5, we discuss the core step of our algorithm: updating the state sequence estimation in response to disordered changes in the observation set. In Section~6, we propose our Gaussian-Bernoulli secure estimator {\color{revision}and analyze its convergence and the error of the resulting state estimate.} In Section~7, we demonstrate our algorithm's performance and compare it with other algorithms. In Section~8, we conclude the paper. For clarity, this paper focuses on the single-sensor system, but the derivation process and conclusions are also applicable to multi-sensor systems. 

\textbf{Notation:} For $S\in R^{n \times n}, S\succ 0$, let $\|x\|_{S} = \sqrt{x^TSx} $ denote the S-weighted 2-norm of $x$. Denote the state sequence at interval $T=[0,N]$ as $x_{0:N}=\{x_0,\cdots,x_N\}$. The observation set is denoted as $y_{0:N}=\{y_0,\cdots,y_N\}$. The observation indicators are denoted as $p_{0:N}=\{p_0,\cdots,p_N\}$. Let $\mathcal{O} \in 2^{\{0,\cdots,N\}}$ denote the selected index set, and then $y_\mathcal{O}$ denotes the selected observation set. $\hat{x}_{i|\mathcal{O}}$ denotes the optimal estimation of the state $x_i$ under the observation set $y_{\mathcal{O}}$. Let the symbol $\circ$ denote the composition of functions.

\section{Bayesian Methods in HMMs}
This section introduces Bayesian estimation as tools for analyzing stochastic systems and estimating the states of a hidden Markov model (HMM). Furthermore, we present the steps of Kalman filter and RTS smoother, highlighting their optimality within our framework.






\subsection{Bayesian Estimation}
Bayesian estimation is a statistical inference method for updating the beliefs in presence of new evidence. This approach is particularly valuable when data is limited or uncertainty is high.  

At the core of Bayesian estimation is Bayes' theorem, which describes the relationship between conditional probabilities. The theorem can be expressed as:
$$p(x|y) = \frac{p(y|x)\cdot p(x)}{p(y)}, $$
where the prior probability $p(x)$ represents current knowledge before observation, the likelihood function $p(y|x)$ describes the known statistical model of observation given the state, and the posterior probability $p(x|y)$ updates probability of the state after observation.

\subsection{MMSE and MAP}
The Minimum Mean Squared Error (MMSE) and Maximum A Posteriori (MAP) are two classic Bayesian estimation methods, each with different probabilistic optimization objectives.

The goal of MMSE estimation is to minimize the mean squared error between the estimated values and the true values. Mathematically, {\color{revision}when the posterior has a finite second moment}, the MMSE estimator can be expressed as:
\begin{align*}
    \hat{x}^{\text{MMSE}} = \mathbf{E}[x|y] = \arg \min_{\hat{x}} \mathbf{E}[
    \| x-\hat{x}\|^2 |y].
\end{align*}
In contrast to the MMSE estimator, the MAP estimator aims to find the state value that maximizes the posterior probability. It means that given the observed data, we seek the state value most likely to produce these data. Mathematically, the MAP estimator can be expressed as:
\begin{align*}
    \hat{x}^{\text{MAP}} = \arg \max_{\hat{x}} p(\hat{x}|y) = \arg \max_{\hat{x}} \frac{p(y|\hat{x})\cdot p(\hat{x})}{p(y)}.
\end{align*}

\subsection{Kalman filter and RTS smoother} \label{sec:C}
The Kalman filter can be considered as a specific application of Bayesian estimation in HMMs. The uniqueness of the Kalman filter is reflected in three main aspects. 

First, both state transitions and observation processes are assumed to be linear and accompanied by Gaussian noise. They can be expressed as follows,
\begin{align*}
    x_{t+1}\mid x_t \sim \mathcal{N}(A{x}_t, Q), \\
    y_t\mid x_t \sim \mathcal{N}(C{x}_t, R),
\end{align*}
where $Q>0$ and $R>0$ denote the process and observation noise covariances. 


{\color{revision}Given the filtering distribution $x_t\mid y_{0:t}\sim\mathcal N(\hat x_{t|t}^*,P_{t|t})$, the observation $y_{t+1}$ updates the estimate of both $x_{t+1}$ and $x_t$. The expectations below are conditional on $y_{0:t}$.}

\textbf{LMMSE}
\begin{equation*} 
\hat{x}^{\text{LMMSE}}_{t+1},K=
\begin{array}{cl} 
 \displaystyle  \arg &\min_{\hat{x}_{t+1},K}  
E[\|x_{t+1} - \hat{x}_{t+1}\|^2]  \\
 \textrm{s.t.} & \hat{x}_{t+1} = A\hat{x}_{t|t}^* +K(y_{t+1}-CA\hat{x}_{t|t}^*)  
\end{array} 
\end{equation*}
\begin{equation*} 
\hat{x}^{\text{LMMSE}}_{t},F=
\begin{array}{clll} 
 \displaystyle  \arg&\min_{\hat{x}_{t},F}  
E[\|x_{t} - \hat{x}_{t}\|^2]  \\
 \textrm{s.t.} &   \hat{x}_{t} = \hat{x}_{t|t}^* +F(\hat{x}^{\text{LMMSE}}_{t+1}-A\hat{x}_{t|t}^*)  
\end{array}
\end{equation*}
\textbf{MAP}
\begin{align*}
&\hat{x}^{\text{MAP}}_{t+1}, \hat{x}^{\text{MAP}}_{t}
= \arg\max_{\hat{x}_{t+1},\hat{x}_{t}} \; p(\hat{x}_{t+1},\hat{x}_{t}\mid y_{0:t+1}) \\
&= \arg\max_{\hat{x}_{t+1},\hat{x}_{t}} \; p(y_{t+1}\mid \hat{x}_{t+1})\,p(\hat{x}_{t+1}\mid \hat{x}_{t})\,p(\hat{x}_{t}\mid y_{0:t}) \\
&= \arg\min_{\hat{x}_{t+1},\hat{x}_{t}} \Big(
\|y_{t+1}-C\hat{x}_{t+1}\|^2_{R^{-1}}
+ \|\hat{x}_{t+1}-A\hat{x}_{t}\|^2_{Q^{-1}}
\\
&+ \|\hat{x}_{t}-\hat{x}_{t|t}^*\|^2_{P_{t|t}^{-1}}
\Big).
\end{align*}

Moreover, LMMSE and MAP have the same result in linear Gaussian model. 
The closed-form solution of optimal estimate can be calculated by Kalman filter and Rauch-Tung-Striebel (RTS) smoother.
\begin{align*}
    \hat{x}_{t+1|t+1} = \hat{x}^{\text{LMMSE}}_{t+1} = \hat{x}^{\text{MAP}}_{t+1} \\
    \hat{x}_{t|t+1} = \hat{x}^{\text{LMMSE}}_{t} = \hat{x}^{\text{MAP}}_{t}
\end{align*}


\section{Problem Formulation}\label{sec:formulation}
This section introduces the Gaussian-Bernoulli mixture observation model and explains its motivation. Following this, we describe our objectives and formulate them as a dual-variable optimization problem involving state estimation and Boolean decision-making. To avoid unnecessary complexity, we describe the problem under a single sensor scenario.

\subsection{Cyber Physical System Model}
We consider a linear time-invariant system with process and measurement noise:
\begin{align}
\left\{\begin{aligned}
 x_{t+1} &= Ax_t + w_t \\
 y_{t} &= Cx_t + \bar{v}_t
\end{aligned}\right.
\end{align}
The state is $x_t\in\mathbb R^n$, the measurement is $y_t\in\mathbb R^m$, and $(A,C)$ is observable.
{\color{revision}We use the centered prior $x_0\sim\mathcal N(0,P_0)$, independent process noises $w_t\sim\mathcal N(0,Q)$ and independent nominal measurement noises $v_t\sim\mathcal N(0,R)$, with mutual independence and $P_0,Q,R\succ0$. A known nonzero prior mean can be removed by centering.}

To model the continuous error term $\bar{v}_t$ together with the possible occurrence of discrete abnormal events, we draw inspiration from mixed probabilistic modeling.
Latent-variable models in biostatistics jointly describe continuous measurements and discrete outcomes~\cite{sammel1997latent}; mixed graphical models couple Gaussian and categorical variables for heterogeneous data~\cite{lee2015learning}; and GaussianFormer-2 uses probabilistic Gaussian representations with occupancy variables for sparse 3D scene modeling~\cite{huang2025gaussianformer}. These works indicate that a single continuous distribution is often inadequate when continuous variations coexist with discrete events or structural switches. Motivated by this insight, we adopt a Gaussian-Bernoulli structure to model observation noise with intermittent deviations.

{\color{revision}
\begin{assumption}[Working observation model]\label{assumption 1}
Let $\bar v_t=v_t+p_t\delta_t$. The detector assigns independent indicators $p_t\sim{\rm Bernoulli}(\beta)$, $0<\beta<1/2$. For both values of $p_t$ it assigns the same independent reference law to $u_t=R^{-1/2}\delta_t$, uniform on the Euclidean ball $\|u_t\|\le B$, with $0<B<\infty$. The indicators and shifts in this working model are independent of the nominal process.
\end{assumption}
Uniform density expresses equal preference among amplitudes within the specified range $B$ in noise-standardized coordinates. Actual deviations can persist across time or be deterministic. 

\subsection{Problem of Interest}\label{sec:profile}
{\normalcolor Our goal is to design a secure detector to determine whether the sequential observations} in $[0,N]$ {\normalcolor are reliable. The secure detector operates in parallel with the estimator that updates the state estimation.} It can therefore use observations before and after $t$ to reassess $y_t$.
For a candidate state let $r_t=R^{-1/2}(y_t-Cx_t)$ and $z_t=\|r_t\|$. We maximize the \emph{joint} density in the latent amplitudes and indicators for the given data. The uniform reference density is common to the two indicator values and cancels from their joint density ratio. Up to a common additive constant, minus twice the joint log density for one observation, after profiling, is
\begin{align}
 \min_{\|u\|\le B}\bigl\{\|r_t-pu\|^2+\alpha p\bigr\},
 \qquad \alpha=2\log\frac{1-\beta}{\beta}. \label{eq:prior-odds}
\end{align}
For $p=1$ the minimizer is the projection of $r_t$ onto the ball. Optimizing also over $p$ gives
\begin{align}
 \rho_B(z)&=\min\{z^2,\,\alpha+(z-B)_+^2\}, \label{eq:bounded-profile}\\
 \rho_\infty(z)&=\min\{z^2,\alpha\}, \qquad (a)_+=\max(a,0).
 \label{eq:capped-loss}
\end{align}
Integrating the amplitude out instead would give a Gaussian--uniform convolution and a different likelihood. When $\beta$ is unavailable, $\alpha$ can be chosen as a design penalty. The resulting $\hat p_t$ is a binary decision. A sensor alarm counts such decisions over the observation window, so it flags repeated deviations without identifying their cause.
}

\begin{figure}[htbp]
\centering
\includegraphics[width=0.45\textwidth]{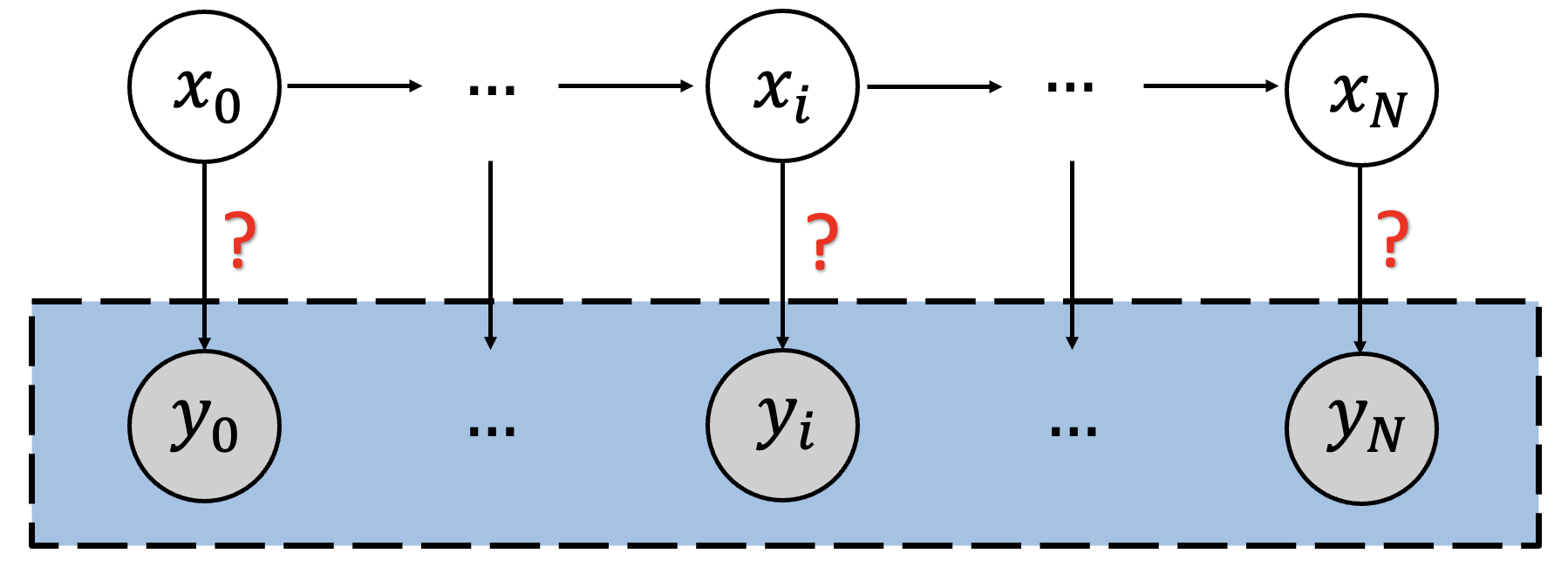}
\caption{Illustration of the detection of sensor attacks.}
\end{figure}

{\color{revision}
For $X=\operatorname{col}(x_0,\ldots,x_N)$, define the nominal trajectory cost
\begin{equation}
 d(X)=\|x_0\|_{P_0^{-1}}^2+
 \sum_{t=1}^N\|x_t-Ax_{t-1}\|_{Q^{-1}}^2 . \label{eq:dyn-prior}
\end{equation}
Proposition~\ref{prop:wide-prior} gives a sufficient radius in terms of the observed batch and the nominal model parameters.

\begin{proposition}[Finite reference range]\label{prop:wide-prior}
Put $n_o=N+1$ and write $d(X)=X^\top H_0X$. Define
\begin{align}
 B_*={}&\max_{0\le t\le N}\|R^{-1/2}y_t\|\notag\\[-2pt]
 &+\|R^{-1/2}C\|_2\sqrt{n_o\alpha/\lambda_{\min}(H_0)} .
 \label{eq:sufficient-radius}
\end{align}
For $B\ge B_*$, the objectives $d(X)+\sum_t\rho_B(z_t)$ and $d(X)+\sum_t\rho_\infty(z_t)$ have the same minimum and the same set of minimizing trajectories. Their minimizing indicator choices also agree at each such trajectory.
\end{proposition}
\textit{Proof.} The prior and the dynamics give $H_0\succ0$, since $d(X)=0$ implies successively $x_0=0$ and $x_t=0$ for all $t$. The wide-range objective at $X=0$ is at most $n_o\alpha$. Every minimizer therefore satisfies $\|X\|\le\sqrt{n_o\alpha/\lambda_{\min}(H_0)}$. Its residuals are all at most $B_*$, so both indicator costs agree there. The finite-range objective dominates the wide-range objective everywhere. Equality on all minimizers of the latter proves both claims. \hfill$\square$

If a smaller radius leaves part of an abnormal residual unexplained, the term $(z-B)_+^2$ can pull the state estimate toward that reading. In the wide-range regime, retaining the indicators explicitly gives

\noindent\textbf{Problem 1.}
\begin{equation}
\begin{aligned}
 &\min_{X,\,p\in\{0,1\}^{n_o}}\mathcal W(X,p),\\
 &\mathcal W(X,p)=d(X)+\sum_{t=0}^{N}
 \big[(1-p_t)z_t^2+\alpha p_t\big].
\end{aligned}\label{problem 1}
\end{equation}
{\normalcolor The problems of obtaining $\{\hat{p}_t\}$ and $\{\hat{x}_t\}$ are coupled.} {\normalcolor It is a Mixed-Integer Programming (MIP) problem with continuous variables ($\hat{x}_{0:N}$) and Boolean variables ($\hat{p}_{0:N}$).} Given the indicators, the state update is a Gaussian smoothing problem on the retained observations. Optimizing over all indicators entails a combinatorial search.
}

\begin{figure}[htbp]
\centering
\includegraphics[width=0.47\textwidth]{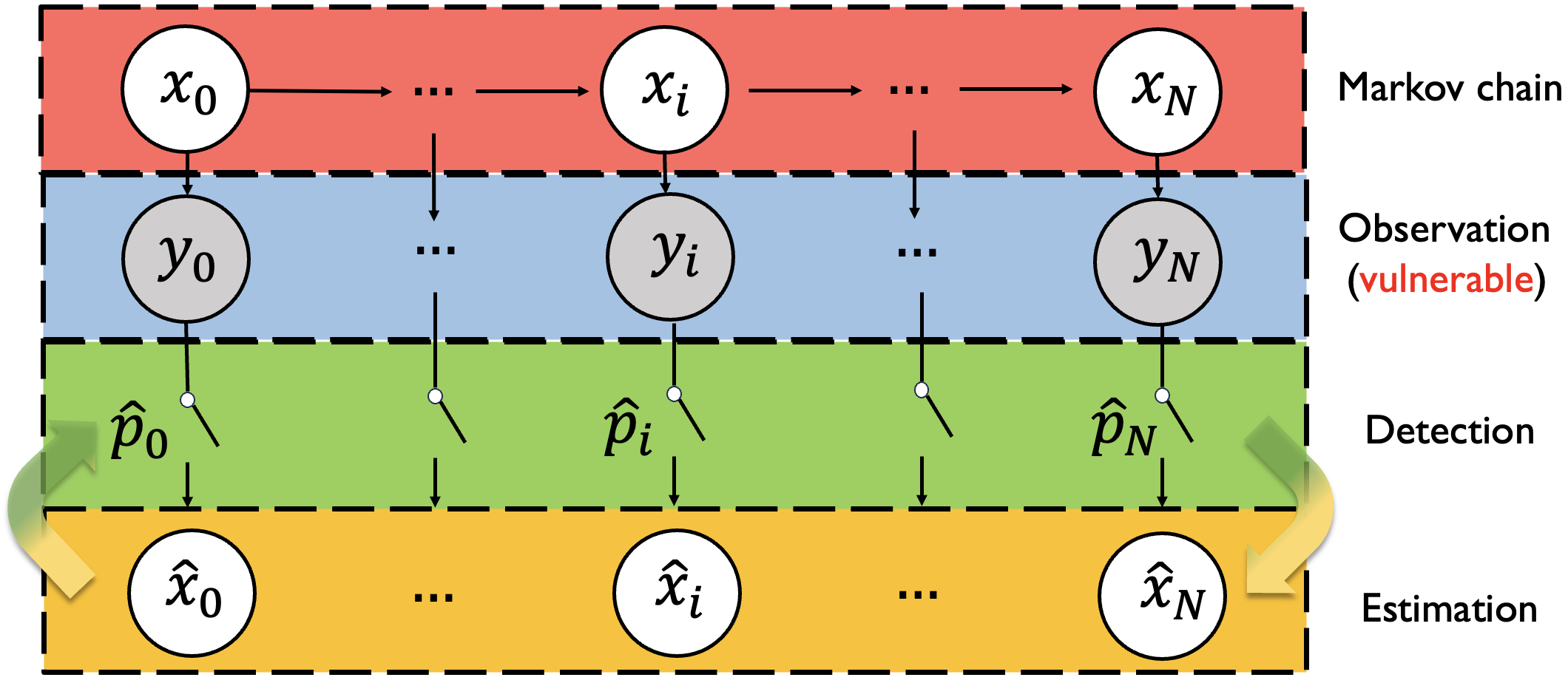}
\caption{\textcolor{revision}{Red represents the system states, blue the observations, green the observation indicators, and yellow the estimates based on retained observations.}}
\end{figure}
\section{State Estimation with Secure Data: A Direct Computational Approach} \label{sec:Direct}
In this section, we first solve a special case of Problem 1, i.e., when the observation indicators are fixed, what is the closed-form solution of equation (\ref{problem 1})? Furthermore, when the observation indicators change, how should the solution of equation (\ref{problem 1}) be updated? These are the two key steps of our algorithm.

When $\hat{p}_{0:N}$ is fixed, we can denote the selected index set as $\mathcal{O}=\{i|\hat{p}_i=0\}$, and then Problem 1 is reduced into the following equivalent problem.

\textbf{Problem 2}
\begin{align}
    \arg\min_{\hat{x}_{0:N}} ~\mathcal{L}_{\mathcal{O}}&(\hat{x}_{0:N})= \Biggl\{\sum_{i \in \mathcal{O}} \|y_i - C\hat{x}_i\|^2_{R^{-1}} \notag \\ 
    + &\sum_{i=1}^N \|\hat{x}_i - A\hat{x}_{i-1}\|^2_{Q^{-1}} 
    + \|\hat{x}_0\|^2_{P_0^{-1}}\Biggr\} \label{equation map}
\end{align} 
The optimization objective of equation (\ref{equation map}) corresponds to a MAP state estimation problem. This problem can be translated into the following question. The goal is to estimate the state sequence $x_{0:N}$ based on a partial observation set $ y_\mathcal{O} = \{y_i \mid i \in \mathcal{O}\}$, where $y_\mathcal{O} \subseteq y_{0:N}$, in which the measurement errors follow a zero-mean Gaussian distribution.

As mentioned in Section \ref{sec:C}, the solutions for MAP and LMMSE estimation in linear Gaussian systems are the same. Therefore, in this section, we present a direct solution to equation (\ref{equation map}) from the perspective of LMMSE.

\subsection{Sequential Estimation with Partial Observations}
Under the condition of partial observations $y_\mathcal{O}$, our goal is to estimate the state sequence of the system. For a linear Gaussian system, we only need to solve an LMMSE problem, which involves finding the optimal linear relationship between the observations and the states. 

\begin{figure}[htbp]
\centering
\includegraphics[width=\columnwidth]{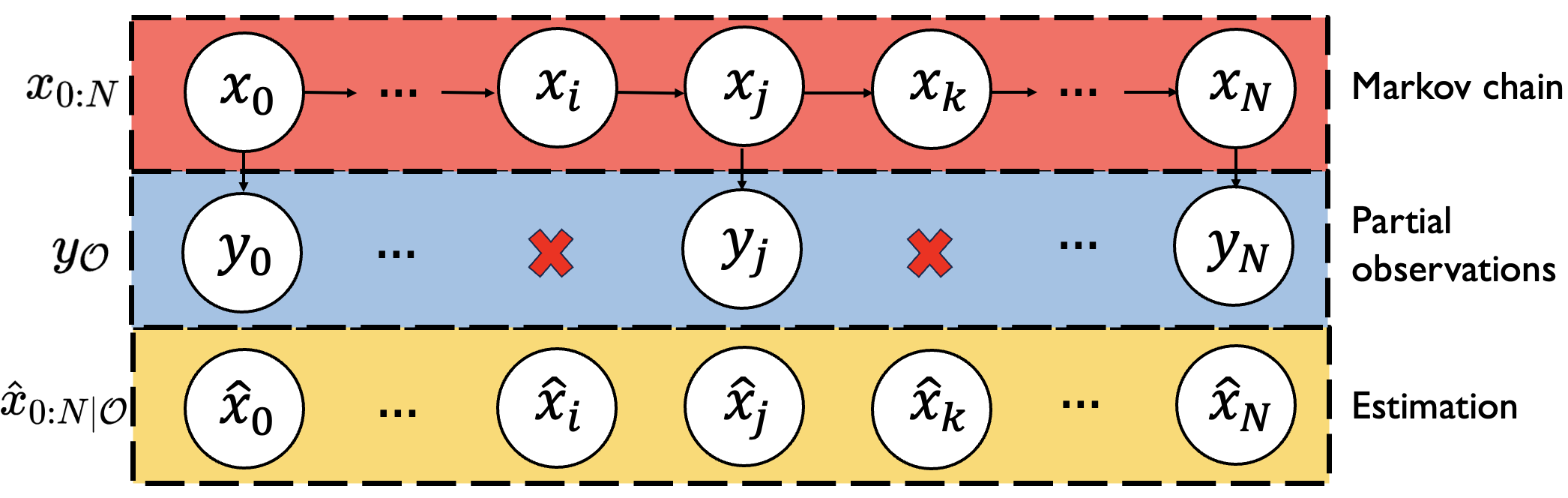}
\caption{The estimation of state sequence under partial observations $y_\mathcal{O}$.}
\end{figure}

Define the column vectors $X = col\{x_0,\cdots,x_N\}, Y = col\{y_0,\cdots, y_N\}$ and $\hat{X}_\mathcal{O} = col\{\hat{x}_{0|\mathcal{O}},\cdots,\hat{x}_{N|\mathcal{O}}\}$. The following formulation describes the sequential estimation problem with partial observations.

{\color{revision}
\noindent\textbf{Problem 3.}
For a prescribed set $\mathcal O$, assume the retained observations
follow the nominal linear Gaussian model in Section~\ref{sec:C}.
Among linear functions of $Y_{\mathcal O}=\operatorname{col}(y_i:i\in\mathcal O)$
find the estimate minimizing $\mathbb E\|X-KY_{\mathcal O}\|_2^2$.
This fixed-set model supplies the fitting rule even when the detector
later selects the set from the data.

\begin{lemma}\label{lemma:1}
Let $P_0,Q,R\succ0$. For each fixed set $\mathcal O$, Problem~2 has a
unique solution. Under the nominal model of Problem~3 it is the Gaussian
conditional mean and can be computed by Kalman filtering with the
measurement update skipped when $i\notin\mathcal O$, followed by one
RTS backward pass. With the zero prior mean used here the solution is
linear in $Y$,
\[
 \hat X_{\mathcal O}=H_{\mathcal O}M_{\mathcal O}L_{\mathcal O}^{-1}Y.
\]
Its block columns for omitted observations vanish.
\end{lemma}
Appendix~\ref{appendices:A} gives the block matrices and the proof.

\begin{remark}
Simultaneous measurements from sensors with independent measurement
noises may be processed in any order at a given time. For correlated
sensor noises the retained measurements are updated jointly with their
corresponding covariance block.
\end{remark}
}

\subsection{Estimation Update with New Observations}
Suppose that we have obtained the optimal estimation under $y_{\mathcal{O}}$. When a new observation $y_k$ is added to the observation set, the optimal estimation of state sequence should be updated with new observation set $y_{\mathcal{O}^*} = y_{\mathcal{O}} \cup \{y_k\}$. {\color{revision}The update uses the nominal covariance $R$ for the added observation.} 
\begin{figure}[htbp]
\centering
\includegraphics[width=\columnwidth]{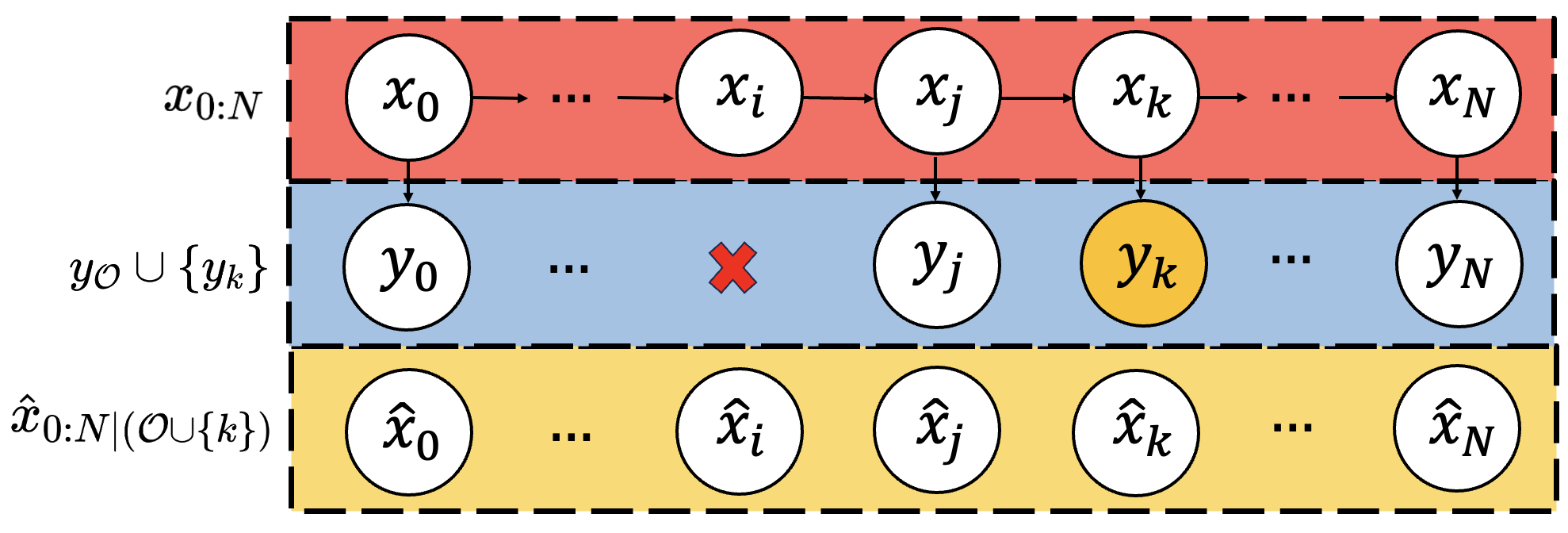}
\caption{Estimation update for new observation $y_k$ under the previous observation set $y_\mathcal{O}$.} \label{switching}
\end{figure}

\begin{figure}[htbp]
\centering
\includegraphics[width=0.3\textwidth]{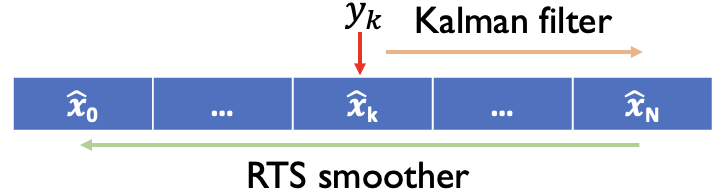}
\caption{Recalculation steps for estimation update.} \label{update step}
\end{figure}

A basic approach is illustrated in Fig.~\ref{update step} by Lemma 1. In detail, we utilize the Kalman gain $\{K_t|t<k\}$ under $y_{\mathcal{O}}$ and recompute the rest process under $y_{\mathcal{O}^*}$.


{\color{revision}Inserting a retained observation at time $k$ changes the filtered gains from $k$ onward and may change smoothed states on the entire window. For a prescribed set of clean measurements, adding a clean observation cannot increase the smoothing error covariance. The next section gives an update initialized at the previous solution.}

\section{State Estimation with Secure Data: An Iterative Computational Approach} 
\label{sec:iter}
In section \ref{sec:Direct}, we discussed how to find the closed-form solution of equation (\ref{equation map}) using a forward and a backward pass. In this section, we propose an iterative method for state sequence estimation that allows for rapid convergence to the optimal estimate. 

The core idea of the iterative method is to fully utilize the previous estimates as the initial guess in the iteration. By constructing a composite convex optimization problem, we design a proximal gradient descent algorithm to avoid redundant computations. {\color{revision}We bound the fixed-set convergence rate and express the initial objective gap in terms of the new innovation.}

\subsection{Sequential Estimation with Partial Observations}
{\color{revision}
For a stacked trajectory $\hat X$, the cost in \eqref{equation map} comprises a measurement term and a trajectory term,
\begin{align}
 \mathcal L_{\rm meas}&=\sum_{i\in\mathcal O}\|y_i-C\hat x_i\|_{R^{-1}}^2,\\
 \mathcal L_{\rm dyn}&=\|\hat x_0\|_{P_0^{-1}}^2+\sum_{i=1}^{N}\|\hat x_i-A\hat x_{i-1}\|_{Q^{-1}}^2.
\end{align}
The block matrix $\tilde{\mathcal A}$ maps $\hat X$ to its prior and transition residuals.
}
\begin{align*}
\begin{bmatrix}
\hat{x}_0 \\
\hat{x}_1 - A \hat{x}_0 \\
\hat{x}_2 - A \hat{x}_1 \\
\vdots \\
\hat{x}_N - A \hat{x}_{N-1}
\end{bmatrix}=
\begin{bmatrix}
I & 0 & 0 & \cdots & 0 \\
-A & I & 0 & \cdots & 0 \\
0 & -A & I & \cdots & 0 \\
\vdots & \vdots & \vdots & \ddots & \vdots \\
0 & \cdots & 0 & -A & I
\end{bmatrix}\begin{bmatrix}
\hat{x}_0 \\
\hat{x}_1 \\
\hat{x}_2 \\
\vdots \\
\hat{x}_N
\end{bmatrix}=
\tilde{\mathcal{A}} \hat{X}.
\end{align*}
The trajectory cost has the quadratic form
\begin{align}
    \mathcal L_{\rm dyn}(\hat X) = \|\tilde{\mathcal{A}}\hat{X}\|_{\tilde{P}^{-1}}^2 = \|\hat{X}\|^2_{\tilde{\mathcal{A}}^T \tilde{P}^{-1} \tilde{\mathcal{A}}} \label{eq:norm}
\end{align}
where $\tilde{P}$ is a block-diagonal matrix,
\begin{align*}
\tilde{P} = 
\begin{bmatrix}
~~P_0~~ & 0 & 0 & \cdots & 0 \\
0 & ~~Q~~ & 0 & \cdots & 0 \\
0 & 0 & ~~Q~~ & \cdots & 0 \\
\vdots & \vdots & \vdots & \ddots & \vdots \\
0 & 0 & 0 & \cdots & ~~Q~~
\end{bmatrix}.
\end{align*}

{\color{revision}
Write $H=\tilde{\mathcal A}^{T}\tilde P^{-1}\tilde{\mathcal A}$ and
let $\tilde C_i$ select $C x_i$ from the stacked trajectory $X$.
For the inner solver we minimize half the cost in Problem~2, so that
the gradients have the usual least-squares normalization.

\noindent\textbf{Problem 4.}
\begin{align}
   \min_X\; &\big\{ f_{\mathcal O}(X)+g(X)\big\},\notag\\
   f_{\mathcal O}(X)&=\tfrac12\sum_{i\in\mathcal O}
     \|y_i-\tilde C_i X\|_{R^{-1}}^2,\notag\\
   g(X)&=\tfrac12\|X\|_H^2. \label{eq:half-cost}
\end{align}
Define $F_{\mathcal O}=\sum_{i\in\mathcal O}\tilde C_i^T R^{-1}\tilde C_i$,
$b_{\mathcal O}=\sum_{i\in\mathcal O}\tilde C_i^T R^{-1}y_i$, and
$B_{\mathcal O}=H+F_{\mathcal O}$. Since $\tilde{\mathcal A}$ is block
triangular with identity diagonal, $H$ is positive definite whenever
$P_0,Q\succ0$. In particular, even the empty observation set has a unique
minimizer. For any $\mathcal O$, it satisfies
\begin{equation}
   B_{\mathcal O}\hat X_{\mathcal O}=b_{\mathcal O}.
   \label{eq:inner-normal}
\end{equation}

\begin{theorem}[Inner rate]\label{theorem:2}
Let $P_0,Q,R\succ0$, $\mu=\lambda_{\min}(H)$, and choose $\eta>0$
with $\eta\lambda_{\max}(F_{\mathcal O})\le1$.
Starting from any $X^0$, the proximal-gradient iteration
\begin{align}
 X^{t+1}=(I+\eta H)^{-1}
 \big[X^t-\eta(F_{\mathcal O}X^t-b_{\mathcal O})\big]
 \label{eq:pg-update}
\end{align}
converges to $\hat X_{\mathcal O}$. If
$E_t=\mathcal L_{\mathcal O}(X^t)-\mathcal L_{\mathcal O}(\hat X_{\mathcal O})$
and $q=(1+\eta\mu)^{-1}$, then
\begin{equation}
 E_t\le q^{2t}E_0,\qquad
 E_t\le \frac{\|B_{\mathcal O}X^t-b_{\mathcal O}\|_2^2}{\mu}.
 \label{eq:inner-rate}
\end{equation}
\end{theorem}
The proof is given in Appendix~\ref{appendices:B}. Choosing $\eta$ for
the full observation set makes it valid for every subset, including
sets of measurements from several sensors. The factorization of $I+\eta H$
can then be reused after the observation set changes.

{\normalcolor We can interpret this iterative process from the perspective of filter and smoother.} The gradient step corrects the trajectory using the residuals of the retained measurements. {\normalcolor The term $(I+\eta H)^{-1}$ can be regarded as a smoother, updating the estimate based on system dynamics.}
}

\subsection{Estimation Update with New Observations}
{\color{revision}
When the set changes to $\mathcal O'$, the previous estimate provides a
starting point for \eqref{eq:pg-update}. The second bound in
\eqref{eq:inner-rate} gives an implementable stopping condition in units
of the original objective.
}

\begin{algorithm}
\small
\caption{Iterative approach for state estimation}\label{alg:2}
{\color{revision}
\begin{algorithmic}[1]
\Require Updated set $\mathcal O'$, previous estimate $X$, tolerance $\delta>0$
\Require $H,\mu,\eta$, and a factorization of $I+\eta H$
\State Form $B_{\mathcal O'}$ and $b_{\mathcal O'}$ as in \eqref{eq:inner-normal}
\While{$\|B_{\mathcal O'} X-b_{\mathcal O'}\|_2^2>\mu\delta$}
    \State Update $X$ once using \eqref{eq:pg-update} with $\mathcal O'$
\EndWhile
\State \Return $X$
\end{algorithmic}}
\end{algorithm}

{\color{revision}
\begin{theorem}[Effect of adding one observation]\label{theorem:3}
Let $i\notin\mathcal O$, $X^0=\hat X_{\mathcal O}$,
$r_i=y_i-\tilde C_i\hat X_{\mathcal O}$, and
$S_i=R+\tilde C_iB_{\mathcal O}^{-1}\tilde C_i^T$. For the updated
set $\mathcal O'=\mathcal O\cup\{i\}$,
\begin{align}
 \hat X_{\mathcal O'}&=\hat X_{\mathcal O}
       +B_{\mathcal O}^{-1}\tilde C_i^T S_i^{-1}r_i,
       \label{eq:trajectory-update}\\
 \min_X\mathcal L_{\mathcal O'}-\min_X\mathcal L_{\mathcal O}
       &=r_i^T S_i^{-1}r_i. \label{eq:profiled-increment}
\end{align}
The initial gap for the updated cost is exactly
\begin{equation}
 E_0=r_i^T(R^{-1}-S_i^{-1})r_i.
 \label{eq:warm-gap}
\end{equation}
For any $\delta>0$, zero iterations suffice if $E_0\le\delta$.
Otherwise $E_t\le\delta$ is guaranteed after
$\lceil\log(E_0/\delta)/(2\log(1/q))\rceil$ iterations, with $q$ from
Theorem~\ref{theorem:2}.
\end{theorem}
Appendix~\ref{appendices:C} proves these identities by completing the
square. Equation~\eqref{eq:warm-gap} relates the work after an insertion
to its innovation and the uncertainty in the predicted measurement. If the added observation equals its smoothed prediction, the estimated trajectory is unchanged. For a fixed previous set, the initial objective gap is quadratic in the innovation.
}
\section{Secure State Estimation Method and Convergence Analysis}
\begin{figure}[htbp]
\centering
\includegraphics[width=0.45\textwidth]{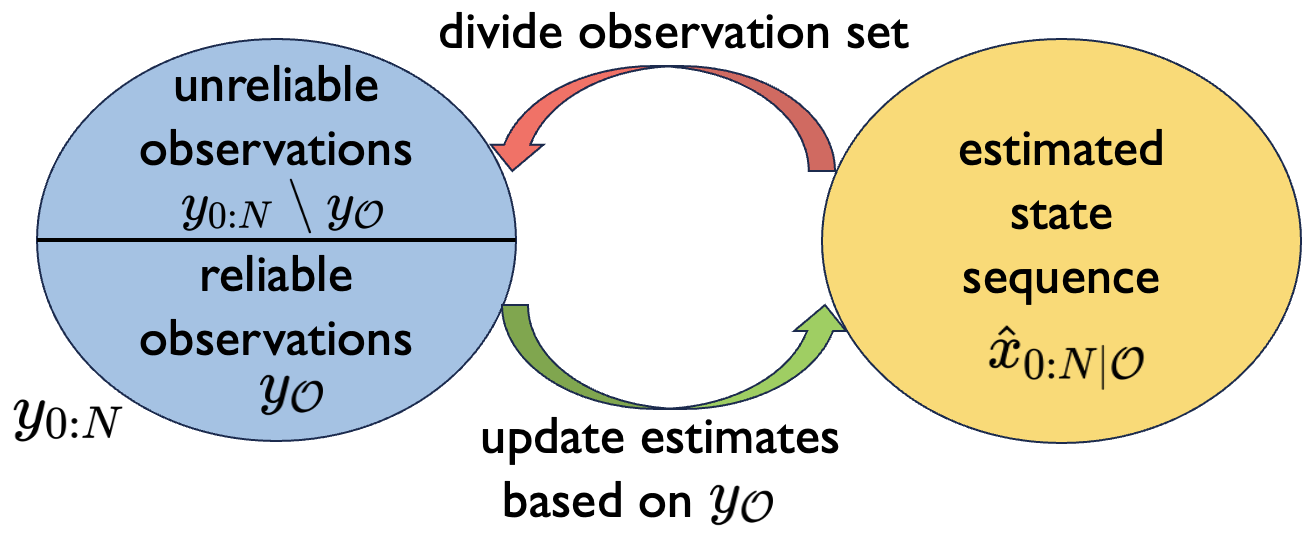}
\caption{The framework of GBS estimator. GBS estimator comprises two components: the resilient estimator and the secure detector. The green arrow indicates the resilient estimator, which estimates the state based on reliable observations. The red arrow denotes the secure detector, which identifies reliable observations based on the estimated state.} \label{MIP solver}
\end{figure}

In this section, we aim to solve the targeted MIP problem presented in \eqref{problem 1}. Summarized in Fig.~\ref{MIP solver}, the GBS estimator involves repeating two steps: assessing the reliability of observations and updating estimates. The algorithms introduced in Section \ref{sec:Direct} and Section  \ref{sec:iter} facilitate the estimation update in response to changes in reliable observation set.

\subsection{Initial Strategy for Rough Estimation}
A good initial value is beneficial for solving problems. However, at first, we do not know the initial values of $\hat{p}_{0:N}$ and $\hat{x}_{0:N}$. Therefore, the initial strategy aims to roughly estimate the value of the observation indicators $p_{0:N}$ based on $y_{0:N}$.

The key idea of the initial strategy is to estimate $\{\hat{x}_t\}$ sequentially and to roughly determine the value of $\hat{p}_{t+1} $ through $\hat{x}_t$. If $\hat{p}_{t+1}=0$, the observation $y_{t+1}$ is utilized for the state estimation of $x_{t+1}$ through Kalman filter. Algorithm~\ref{alg:3} provides the details of initial strategy.

\begin{algorithm}
\small
\caption{Initial strategy for rough estimation}
\label{alg:3}
\begin{algorithmic}[1]
\Require system parameters $\{A, C, Q, R, \alpha\}$
\Require initial state estimate $(\bar{x}_0 = 0, \bar{P}_0)$ and full observation set $y_{0:N}$
\Ensure estimated indicators $\hat{p}_{0:N}$ and states $\hat{x}_{0:N}$

\State Initialize $\hat{x}_0 \gets \bar{x}_0$, $P_0 \gets \bar{P}_0$
\For{$i = 0$ to $N$}
    \State {\color{revision}$\hat{p}_i \gets \mathbf 1_{\{\|y_i-C\hat{x}_i\|^2_{R^{-1}}>\alpha\}}$}
    \If{{\color{revision}$\hat{p}_i=0$}}
        \State $K_i \gets P_i C^T (C P_i C^T + R)^{-1}$
        \State $\hat{x}_i \gets \hat{x}_i + K_i (y_i - C\hat{x}_i)$
        \State $P_i \gets (I - K_i C) P_i$
    \EndIf

    \State $\hat{x}_{i+1} \gets A \hat{x}_i$
    \State $P_{i+1} \gets A P_i A^T + Q$
\EndFor
\end{algorithmic}
\end{algorithm}

\subsection{Gaussian-Bernoulli Secure (GBS) Estimator}
{\color{revision}
Let $\mathcal I=\{0,\ldots,N\}$, $n_o=N+1$, and
$\mathcal O=\{i\in\mathcal I:\hat p_i=0\}$ denote the retained
observations. For fixed indicators, the fitted trajectory and the
value of Problem~1 after fitting are, respectively,
\begin{align}
 \hat X_{\mathcal O}&=\arg\min_X\mathcal L_{\mathcal O}(X),
        \label{eq:17}\\
 J(\mathcal O)&=\mathcal L_{\mathcal O}(\hat X_{\mathcal O})
                 +\alpha(n_o-|\mathcal O|).
        \label{eq:reduced_objective}
\end{align}
For fixed $X$, write $d_i(X)=\|y_i-\tilde C_iX\|_{R^{-1}}^2$.
Retaining $i$ costs $d_i(X)$ and rejecting it costs $\alpha$. Thus
the minimizing update for the retained set is
\begin{equation}
 \mathcal T(X,\mathcal O)
 =\{i:d_i(X)<\alpha\}\cup\{i\in\mathcal O:d_i(X)=\alpha\}.
 \label{eq:18}
\end{equation}
Keeping the current indicator at equality prevents cost-neutral
moves. This simultaneous update holds $X$ fixed. An observation can have a small residual partly because it helped fit the trajectory. We therefore optimize its indicator together with the state sequence, comparing the refitted costs with and without that observation.

For index $i$, put $\mathcal O^-=\mathcal O\setminus\{i\}$,
$\mathcal O^+=\mathcal O^-\cup\{i\}$, and let
$D_i=\min_X\mathcal L_{\mathcal O^+}(X)-\min_X\mathcal L_{\mathcal O^-}(X)$.
By \eqref{eq:profiled-increment}, $D_i$ is the squared innovation under
$\mathcal O^-$ weighted by the inverse innovation covariance. Since
$J(\mathcal O^+)-J(\mathcal O^-)=D_i-\alpha$, the decrease from flipping
the current indicator is
\begin{equation}
 h_i(\mathcal O)=J(\mathcal O)-J(\mathcal O\triangle\{i\})
 =\begin{cases}D_i-\alpha,&i\in\mathcal O,\\
                \alpha-D_i,&i\notin\mathcal O.
   \end{cases}
 \label{eq:20}
\end{equation}
Here $\triangle$ denotes symmetric difference. Algorithm~\ref{alg:4}
first tries the simultaneous update. If its decrease is at most
$\varepsilon$, it tests single flips in chronological order.
The default $\varepsilon=0$ accepts every strict improvement.
}

\begin{algorithm}
\small
\caption{Gaussian-Bernoulli Secure Estimator}\label{alg:4}
{\color{revision}
\begin{algorithmic}[1]
\Require Model, data $y_{0:N}$, $\alpha>0$, alarm threshold $\tau$, tolerance $\varepsilon\ge0$
\State Initialize $\mathcal O=\{i:\hat p_i^0=0\}$ using Algorithm~\ref{alg:3}
\Loop
    \State Compute $\hat X_{\mathcal O}$ from \eqref{eq:inner-normal}
    \State $\mathcal U\gets\mathcal T(\hat X_{\mathcal O},\mathcal O)$
    \If{$J(\mathcal O)-J(\mathcal U)\le\varepsilon$}
        \State $\mathcal U\gets\mathcal O$
        \For{$i=0,\ldots,N$}
            \If{$h_i(\mathcal O)>\varepsilon$}
                \State $\mathcal U\gets\mathcal O\triangle\{i\}$
                \State \textbf{break}
            \EndIf
        \EndFor
    \EndIf
    \If{$\mathcal U=\mathcal O$}
        \State \textbf{break}
    \EndIf
    \State $\mathcal O\gets\mathcal U$
\EndLoop
\State Return $\hat X_{\mathcal O}$ and $\hat p_i=\mathbf 1_{\{i\notin\mathcal O\}}$
\Statex \hspace{\algorithmicindent}with alarm flag $\mathbf 1_{\{n_o-|\mathcal O|>\tau\}}$.
\end{algorithmic}}
\end{algorithm}

\subsection{Convergence Analysis}\label{sec:outer-convergence}
{\color{revision}
The rate in Theorem~\ref{theorem:2} concerns a fixed set. For the
joint problem the natural termination criterion is the largest decrease
available by changing one indicator and refitting the entire trajectory,
\begin{equation}
 G(\mathcal O)=\max\{0,\ \max_{i\in\mathcal I}h_i(\mathcal O)\}.
 \label{eq:stationarity-certificate}
\end{equation}

\begin{theorem}[Outer bound]
\label{theorem:convergence}
Suppose $P_0,Q,R\succ0$, $\alpha>0$, and Algorithm~\ref{alg:4}
uses the exact fixed-set minima. For every initial set $\mathcal O^0$
and every $\varepsilon\ge0$, it returns a set $\mathcal O^\star$ with
$G(\mathcal O^\star)\le\varepsilon$.
Let $J_0=J(\mathcal O^0)$ and
$s_0=\min\{n_o,\lfloor J_0/\alpha\rfloor\}$. The number $K$ of
accepted outer iterations satisfies
\begin{align}
 K&\le\sum_{s=0}^{s_0}\binom{n_o}{s}-1, \label{eq:bound}\\
 K&\le\lfloor J_0/\varepsilon\rfloor,\qquad \varepsilon>0.
 \label{eq:tolerance-complexity}
\end{align}
At $\varepsilon=0$, the trajectory is the unique optimum for the
returned set, and refitting after any single addition or deletion
cannot lower the joint objective.
\end{theorem}

\emph{Proof.}
For the simultaneous proposal $\mathcal U$ and $X=\hat X_{\mathcal O}$,
each changed indicator saves $|d_i(X)-\alpha|$ at fixed $X$. Refitting
can only increase this saving, hence
\begin{equation}
 J(\mathcal O)-J(\mathcal U)
 \ge \sum_{i\in\mathcal U\triangle\mathcal O}|d_i(X)-\alpha|.
 \label{eq:batch-descent}
\end{equation}
A change of a single indicator saves exactly $h_i(\mathcal O)$ by
\eqref{eq:20}. Both acceptance tests therefore give strict descent.
No set repeats and every visited set satisfies
$\alpha(n_o-|\mathcal O|)\le J(\mathcal O)\le J_0$.
Counting the possible rejected sets yields \eqref{eq:bound}. For
$\varepsilon>0$, summing the decreases, each greater than
$\varepsilon$, and using $J\ge0$ yields \eqref{eq:tolerance-complexity}.
At termination the algorithm has tested all single flips at the current
set, so \eqref{eq:stationarity-certificate} is at most $\varepsilon$.
Uniqueness follows from $B_{\mathcal O}\succ0$. \hfill$\square$

The bound counts accepted changes of set. There is one additional
pass to certify termination. The exact bound can grow exponentially
with the window length. The tolerance bound gives an
$O(J_0/\varepsilon)$ worst-case iteration count for a prescribed
one-flip certificate. Joint changes of several indicators can still
improve a certified solution. In particular, the bounds do not measure
the gap to the globally optimal mask. A direct solve of
\eqref{eq:inner-normal} can be used for the outer comparisons. If
Algorithm~\ref{alg:2} supplies approximate objectives instead, its
residual bound in \eqref{eq:inner-rate} bounds their error; any unresolved
comparison can be settled with the direct solve.
}

{\color{revision}
\subsection{Stability under observation selection}
\label{sec:selected-stability}

The set $\mathcal O$ changes both with the data and during the search.
For each realized set, let $\hat X=(\hat x_0,\ldots,\hat x_N)$ be the
exact minimizer of the state subproblem and $X=(x_0,\ldots,x_N)$ the
true trajectory, with the blocks stacked as columns. Write
$\mathcal G$ for the genuinely clean observations retained in
$\mathcal O$ and $\mathcal B=\mathcal O\setminus\mathcal G$.
These sets are used only for analysis. Put $C_R=R^{-1/2}C$, $c=\|C_R\|_2$,
and define the prior and dynamics operator by
\[
 (Dh)_j=\begin{cases}
 P_0^{-1/2}h_0, &j=0,\\
 Q^{-1/2}(h_j-Ah_{j-1}), &j\ge1.
 \end{cases}
\]
For $I=\{s,\ldots,s+L-1\}$, consider the condition
\begin{equation}
 \sum_{j\in I\cap\mathcal G}(A^{j-s})^\top C_R^\top C_R A^{j-s}
 \succeq \gamma I_n .
 \label{eq:clean-window}
\end{equation}
It requires the clean observations actually used in each window to
distinguish all state directions. For the example $A=C=I_n$, one
retained clean vector measurement per window suffices.

\begin{theorem}[Selected-set error bound]
\label{thm:selected-stability}
Suppose $P_0,Q,R\succ0$. Define
\[
 \begin{split}
 z_j&=\|(D^\top D X)_j\|_2+c\|R^{-1/2}v_j\|_2,\\
 b_j&=\mathbf{1}_{\{j\in\mathcal B\}}\,
        c\,\|R^{-1/2}(y_j-C\hat x_j)\|_2 .
 \end{split}
\]
On each fixed horizon there are $\mu>0$ and $0<\rho<1$, common to
all selected sets, such that at every exact state update
\begin{equation}
 \|\hat x_t-x_t\|_2\le
 \frac{1}{\mu}\sum_{j=0}^N\rho^{|t-j|}(z_j+b_j).
 \label{eq:selected-decay}
\end{equation}
If \eqref{eq:clean-window} holds in every window for fixed integer
$L\ge1$ and $\gamma>0$
and $N+1\ge L$, the same constants can be chosen independently of $N$.
At a terminal GBS solution with exact one-flip optimality, $b_j\le c\sqrt{\alpha}$ for all $j$.
Thus, if the window condition holds almost surely across the horizons
considered and $s_z=\sup_{N,j}(\mathbb E z_j^2)^{1/2}<\infty$, then
\begin{equation}
 \sup_{N,t}(\mathbb E\|\hat x_t-x_t\|_2^2)^{1/2}
 \le \frac{1+\rho}{\mu(1-\rho)}(s_z+c\sqrt{\alpha}).
 \label{eq:selected-rms}
\end{equation}
\end{theorem}

Appendix~\ref{app:selected-stability} proves the result using the
state optimality equations for the realized set. Because selection uses
the measurements, the nominal covariance for a prescribed mask need
not equal the actual error covariance after selection. The bound covers
errors in attack identification and permits arbitrary dependence
between the attack values, the nominal noises and the selection rule
whenever the stated conditions hold. A retained corrupted reading
enters the final bound through its residual at the fitted trajectory.
For a one-flip tolerance $\varepsilon>0$, the same argument replaces $\sqrt{\alpha}$ by $\sqrt{\alpha+\varepsilon}$. A discarded clean reading can reduce the information in
\eqref{eq:clean-window}. Under the stated noise model, $s_z$ is finite
because each block of $D^\top D X$ uses only the prior error and/or
adjacent process innovations.

For a fixed number of sensors with independent nominal errors, let $C_{R,\ell}=R_\ell^{-1/2}C_\ell$ and $c_\ell=\|C_{R,\ell}\|_2$. Sum the Gramian in \eqref{eq:clean-window} over the clean retained sensor--time pairs in the window. In the proof, the nominal noise block is bounded by $\sum_\ell c_\ell\|R_\ell^{-1/2}v_{j,\ell}\|_2$, and the terminal corrupted block by $\sqrt{\alpha}\sum_\ell c_\ell$. The same time-block bandwidth and decay argument apply. Thus one retained clean sensor in each window suffices in the scalar example with $A=C_\ell=1$.

Full-system observability alone does not ensure a bound uniform in
time for every mask. Consider the observable scalar random walk
$A=C=1$ with a prescribed empty retained set. The minimizer is
$\hat x_t=0$ and $\mathbb E|\hat x_t-x_t|^2=P_0+tQ$.
}
\section{Numerical Simulations}
In this section, we introduce the comprehensive experimental results. First, we compare the update speeds of direct and iterative methods under different observation update rules. Then, we compare our GBS estimator with other detectors and estimators from the perspectives of detection and estimation performance. Next, we demonstrate the detection capabilities of our GBS estimator in a multi-sensor system. 

Although our previous focus is on the single-sensor system, the derivation process and conclusions are also applicable to multi-sensor systems. For a multi-sensor system equipped with $M$ sensors, $y_{i,j}$ denotes the observation at the sensor $j$ at time $i$, and the corresponding indicator is defined as $p_{i,j}$.

\subsection{Comparison of Direct and Iterative Approach}
Consider a 3-dimensional system with $M$ sensors over the interval $T = [0,N]$:
\begin{align*}
    M=N=100, \hspace{0.2cm} x_0 \sim \mathcal{N}(0,I), \hspace{0.2cm} A=C=I, \\
    Q_i = R_j =\begin{bmatrix}
    1  & 0.5 & 0\\
     0.5 & 1 & 0.5\\
     0 & 0.5 & 1
    \end{bmatrix}, \begin{matrix}
 i= 0,...,N\\
j=1,...,M
\end{matrix}.
\end{align*}
The full observation set is $y_{0:N,1:M}$. The observation set is updated according to a specific pattern, denoted as $\mathcal{O}_k \in 2^{\{0,\cdots,N\}\otimes \{1,\cdots,M\}}$ for the $k-$th update. Our algorithm is to update the state estimates of $x_{0:N}$ based on the observation set $\mathcal{O}_k$.

\begin{figure}[htbp]
    \begin{subfigure}[b]{\columnwidth}
        \begin{minipage}{0.52\linewidth}
            \includegraphics[width=\linewidth]{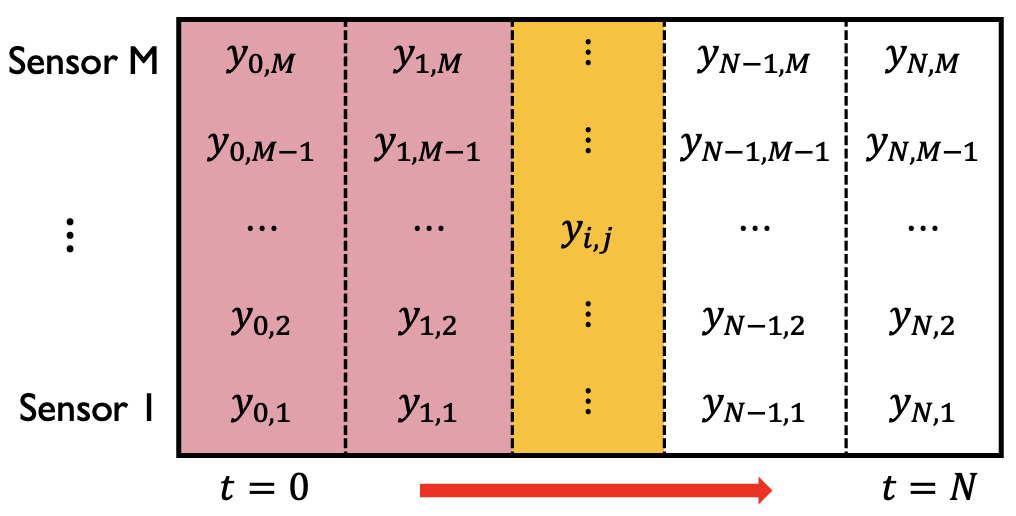}
        \end{minipage}
        \begin{minipage}{0.40\linewidth}
            \includegraphics[width=\linewidth]{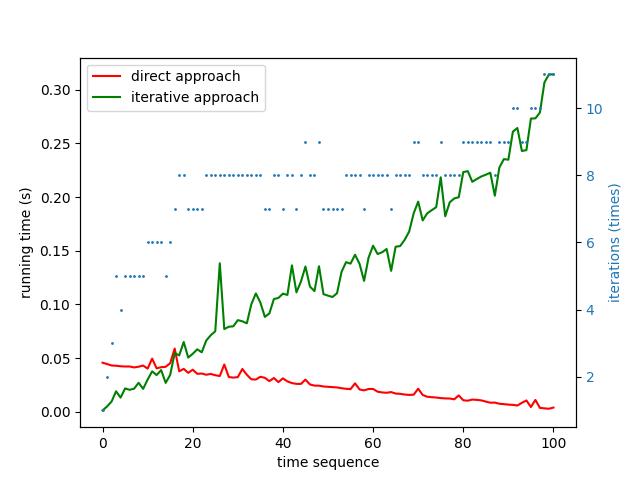}
        \end{minipage}
        \caption{add observations along the timeline}     \label{add observations}
    \end{subfigure}
    \begin{subfigure}[b]{\columnwidth}
        \begin{minipage}{0.52\linewidth}
            \includegraphics[width=\linewidth]{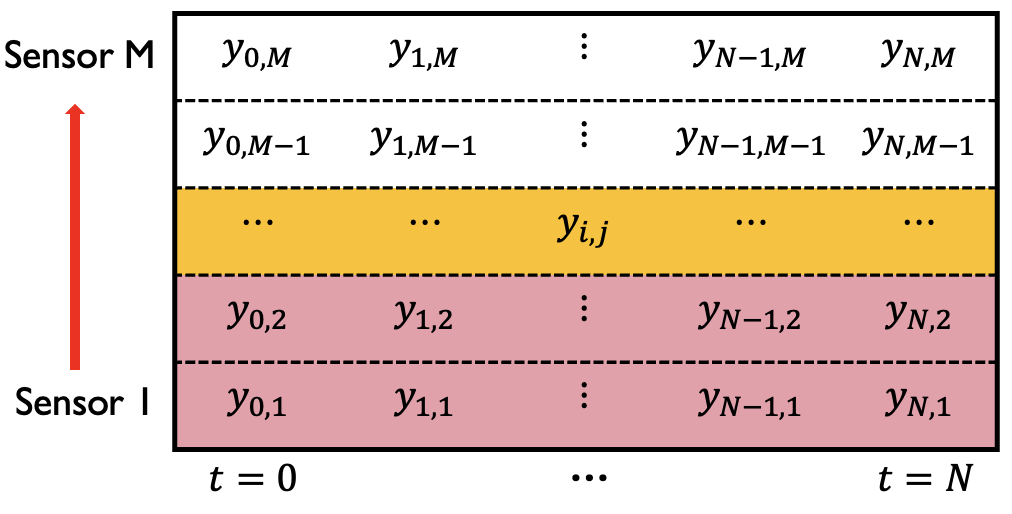}
        \end{minipage}
        \begin{minipage}{0.40\linewidth}
            \includegraphics[width=\linewidth]{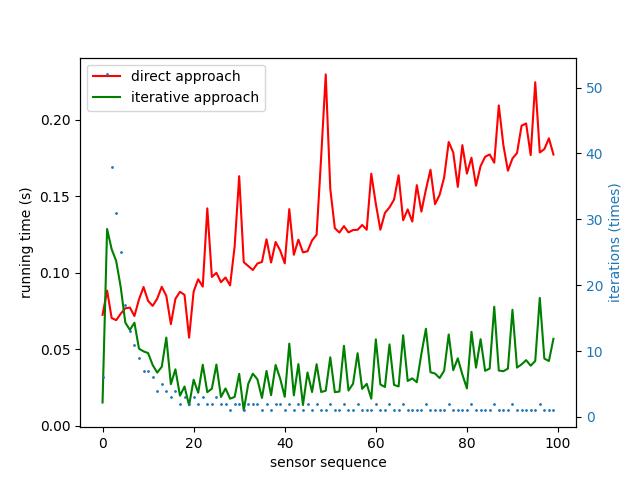}
        \end{minipage}
        \caption{add observations in the order of sensor indices}     \label{add sensor}
    \end{subfigure}
    \begin{subfigure}[b]{\columnwidth}
        \begin{minipage}{0.52\linewidth}
            \includegraphics[width=\linewidth]{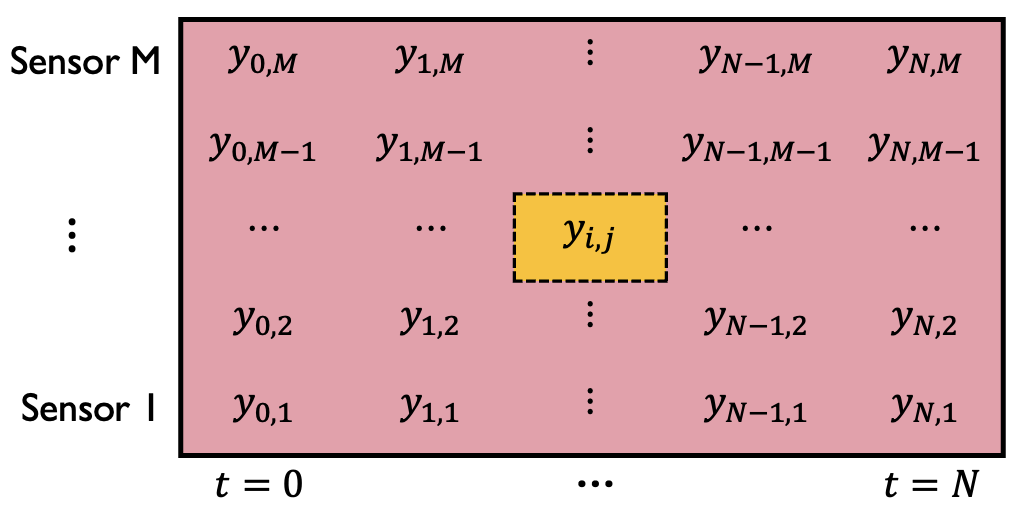}
        \end{minipage}
        \begin{minipage}{0.40\linewidth}
            \includegraphics[width=\linewidth]{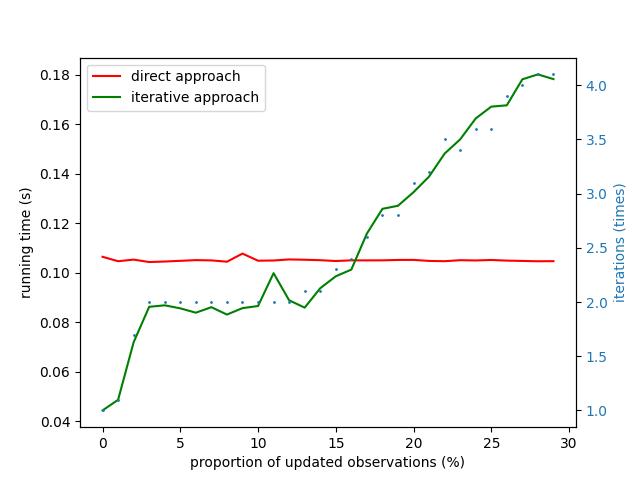}
        \end{minipage}
        \caption{insert the different proportions of observations}     \label{insert observations}
    \end{subfigure}
    \caption{Comparison of direct approach and iterative approach under different observation update rules. In the left figure, the arrow indicates the direction of the observation updates, with the pink representing the past observation set and the yellow representing the newly added observations. In the right figure, the blue dots represent the number of iterations of the iterative approach.}
\end{figure}

First, in Fig.~\ref{add observations}, we implement experiments to add observations along the timeline, which is the default observation update rule for most systems. The newly added observations include the observations from all sensors at time $k$ that $\mathcal{O}_k \setminus \mathcal{O}_{k-1} = \{k\} \otimes \{1,\cdots,M\}$.  Additionally, in Fig.~\ref{add sensor}, we are interested in incorporating observations in the order of sensor indices. The rules for observation updates allow the system to update one after another based on the reliability of the sensors. The new observations add to the set from $k$-th sensor that $\mathcal{O}_k \setminus \mathcal{O}_{k-1} = \{0,\cdots,N\} \otimes \{k\}$. Moreover, in Fig.~\ref{insert observations}, we test randomly updated observation sets, similar to the setup in Algorithm\ref{alg:4}. We first hide part of the observations and compute the optimal estimate. Then, we add the masked observations back and record the computation time required to update the estimate.

The results show that the direct and iterative methods work best in different situations. The direct method is more efficient for timeline-based updates, while the iterative method is better when a small, unordered part of the observation set is updated. The iterative method is also preferred when adding new sensors.

\subsection{Comparison with Other Detectors and Estimators}\label{sec:paired}
\begin{table*}[!t]
    \centering
    \caption{\textcolor{revision}{Detection and estimation results over 10,000 trials per setting. In each column, darker shading marks a better rank among the five methods (higher success or lower error), and tied values share a shade.}}\label{tab:sensor-performance}
    \small
    \renewcommand{\arraystretch}{1.15}
    \setlength{\tabcolsep}{.65pt}
    \setlength{\fboxsep}{1.6pt}
    \definecolor{detectionTint}{HTML}{8DB6CA}
\definecolor{errorTint}{HTML}{D7A47D}
\newcommand{\gbsNumber}[1]{\makebox[3.15em][c]{\rule[-.12ex]{0pt}{1.95ex}#1}}
\newcommand{\gbsCell}[3]{\colorbox{#1!#2}{\gbsNumber{#3}}}
\begin{tabular}{@{}l@{\hspace{.9em}}c@{\hspace{1.15em}}*{5}{c}@{\hspace{1.15em}}*{5}{c}@{}}
\toprule
 & No attack & \multicolumn{5}{c}{\bfseries Independent interference (variance $v$)} & \multicolumn{5}{c}{\bfseries Sensor replay (lag $T$ in steps)} \\
\cmidrule(lr){2-2}\cmidrule(lr){3-7}\cmidrule(lr){8-12}
Parameter value & 0 & 2 & 4 & 6 & 8 & 10 & 2 & 4 & 6 & 8 & 10 \\
\addlinespace[3pt]
\multicolumn{12}{@{}l}{\bfseries\itshape Detection success (\%)} \\
\hspace{.6em}GBS & \gbsCell{detectionTint}{56}{98.58} & \gbsCell{detectionTint}{100}{52.79} & \gbsCell{detectionTint}{100}{55.88} & \gbsCell{detectionTint}{100}{58.69} & \gbsCell{detectionTint}{100}{61.05} & \gbsCell{detectionTint}{100}{63.11} & \gbsCell{detectionTint}{34}{73.94} & \gbsCell{detectionTint}{100}{82.60} & \gbsCell{detectionTint}{100}{93.28} & \gbsCell{detectionTint}{78}{97.14} & \gbsCell{detectionTint}{100}{98.70} \\
\noalign{\vskip .65pt}
\hspace{.6em}$\chi^2$ & \gbsCell{detectionTint}{56}{98.58} & \gbsCell{detectionTint}{56}{52.59} & \gbsCell{detectionTint}{78}{55.55} & \gbsCell{detectionTint}{78}{58.24} & \gbsCell{detectionTint}{78}{60.54} & \gbsCell{detectionTint}{78}{62.57} & \gbsCell{detectionTint}{56}{73.99} & \gbsCell{detectionTint}{56}{78.41} & \gbsCell{detectionTint}{56}{85.62} & \gbsCell{detectionTint}{56}{91.25} & \gbsCell{detectionTint}{56}{93.38} \\
\noalign{\vskip .65pt}
\hspace{.6em}CUSUM & \gbsCell{detectionTint}{100}{99.95} & \gbsCell{detectionTint}{12}{50.16} & \gbsCell{detectionTint}{12}{50.54} & \gbsCell{detectionTint}{12}{51.09} & \gbsCell{detectionTint}{12}{51.70} & \gbsCell{detectionTint}{12}{52.36} & \gbsCell{detectionTint}{100}{74.73} & \gbsCell{detectionTint}{34}{77.39} & \gbsCell{detectionTint}{12}{79.54} & \gbsCell{detectionTint}{12}{81.20} & \gbsCell{detectionTint}{12}{82.38} \\
\noalign{\vskip .65pt}
\hspace{.6em}Resilient & \gbsCell{detectionTint}{34}{98.54} & \gbsCell{detectionTint}{78}{52.60} & \gbsCell{detectionTint}{56}{55.54} & \gbsCell{detectionTint}{56}{58.22} & \gbsCell{detectionTint}{56}{60.50} & \gbsCell{detectionTint}{56}{62.51} & \gbsCell{detectionTint}{78}{74.10} & \gbsCell{detectionTint}{78}{80.69} & \gbsCell{detectionTint}{78}{92.38} & \gbsCell{detectionTint}{100}{97.44} & \gbsCell{detectionTint}{78}{98.46} \\
\noalign{\vskip .65pt}
\hspace{.6em}Sparse-offset & \gbsCell{detectionTint}{78}{99.23} & \gbsCell{detectionTint}{34}{52.11} & \gbsCell{detectionTint}{34}{54.82} & \gbsCell{detectionTint}{34}{57.35} & \gbsCell{detectionTint}{34}{59.62} & \gbsCell{detectionTint}{34}{61.61} & \gbsCell{detectionTint}{12}{73.87} & \gbsCell{detectionTint}{12}{76.27} & \gbsCell{detectionTint}{34}{81.43} & \gbsCell{detectionTint}{34}{87.05} & \gbsCell{detectionTint}{34}{90.33} \\
\noalign{\vskip .65pt}
\addlinespace[3pt]
\multicolumn{12}{@{}l}{\bfseries\itshape Mean state estimation error} \\
\hspace{.6em}GBS & \gbsCell{errorTint}{78}{0.145} & \gbsCell{errorTint}{78}{0.164} & \gbsCell{errorTint}{78}{0.175} & \gbsCell{errorTint}{78}{0.182} & \gbsCell{errorTint}{78}{0.186} & \gbsCell{errorTint}{100}{0.189} & \gbsCell{errorTint}{100}{0.208} & \gbsCell{errorTint}{100}{0.209} & \gbsCell{errorTint}{100}{0.178} & \gbsCell{errorTint}{100}{0.164} & \gbsCell{errorTint}{100}{0.159} \\
\noalign{\vskip .65pt}
\hspace{.6em}$\chi^2$ / CUSUM & \gbsCell{errorTint}{56}{0.155} & \gbsCell{errorTint}{56}{0.174} & \gbsCell{errorTint}{56}{0.190} & \gbsCell{errorTint}{34}{0.205} & \gbsCell{errorTint}{34}{0.219} & \gbsCell{errorTint}{34}{0.232} & \gbsCell{errorTint}{56}{0.210} & \gbsCell{errorTint}{56}{0.295} & \gbsCell{errorTint}{56}{0.391} & \gbsCell{errorTint}{34}{0.485} & \gbsCell{errorTint}{34}{0.569} \\
\noalign{\vskip .65pt}
\hspace{.6em}Resilient & \gbsCell{errorTint}{34}{0.161} & \gbsCell{errorTint}{34}{0.179} & \gbsCell{errorTint}{34}{0.191} & \gbsCell{errorTint}{56}{0.197} & \gbsCell{errorTint}{56}{0.202} & \gbsCell{errorTint}{56}{0.204} & \gbsCell{errorTint}{100}{0.208} & \gbsCell{errorTint}{78}{0.234} & \gbsCell{errorTint}{78}{0.197} & \gbsCell{errorTint}{78}{0.178} & \gbsCell{errorTint}{78}{0.175} \\
\noalign{\vskip .65pt}
\hspace{.6em}Sparse-offset & \gbsCell{errorTint}{100}{0.140} & \gbsCell{errorTint}{100}{0.156} & \gbsCell{errorTint}{100}{0.168} & \gbsCell{errorTint}{100}{0.178} & \gbsCell{errorTint}{100}{0.185} & \gbsCell{errorTint}{78}{0.192} & \gbsCell{errorTint}{78}{0.209} & \gbsCell{errorTint}{34}{0.320} & \gbsCell{errorTint}{34}{0.417} & \gbsCell{errorTint}{56}{0.477} & \gbsCell{errorTint}{56}{0.502} \\
\noalign{\vskip .65pt}
\addlinespace[3pt]
\midrule
\addlinespace[2pt]
 & No attack & \multicolumn{5}{c}{\bfseries Constant bias} & \multicolumn{5}{c}{\bfseries Increasing bias} \\
\cmidrule(lr){2-2}\cmidrule(lr){3-7}\cmidrule(lr){8-12}
Peak bias $h$ & 0 & 2 & 4 & 6 & 8 & 10 & 2 & 4 & 6 & 8 & 10 \\
\addlinespace[3pt]
\multicolumn{12}{@{}l}{\bfseries\itshape Detection success (\%)} \\
\hspace{.6em}GBS & \gbsCell{detectionTint}{56}{98.58} & \gbsCell{detectionTint}{34}{74.01} & \gbsCell{detectionTint}{100}{84.38} & \gbsCell{detectionTint}{100}{95.75} & \gbsCell{detectionTint}{100}{99.19} & \gbsCell{detectionTint}{100}{99.87} & \gbsCell{detectionTint}{100}{53.69} & \gbsCell{detectionTint}{100}{62.55} & \gbsCell{detectionTint}{100}{70.85} & \gbsCell{detectionTint}{100}{76.88} & \gbsCell{detectionTint}{100}{80.99} \\
\noalign{\vskip .65pt}
\hspace{.6em}$\chi^2$ & \gbsCell{detectionTint}{56}{98.58} & \gbsCell{detectionTint}{56}{74.39} & \gbsCell{detectionTint}{56}{79.65} & \gbsCell{detectionTint}{56}{87.70} & \gbsCell{detectionTint}{56}{92.78} & \gbsCell{detectionTint}{56}{93.25} & \gbsCell{detectionTint}{56}{53.05} & \gbsCell{detectionTint}{56}{55.88} & \gbsCell{detectionTint}{56}{61.38} & \gbsCell{detectionTint}{56}{67.75} & \gbsCell{detectionTint}{56}{72.66} \\
\noalign{\vskip .65pt}
\hspace{.6em}CUSUM & \gbsCell{detectionTint}{100}{99.95} & \gbsCell{detectionTint}{100}{74.96} & \gbsCell{detectionTint}{34}{77.87} & \gbsCell{detectionTint}{12}{80.14} & \gbsCell{detectionTint}{12}{81.91} & \gbsCell{detectionTint}{12}{83.19} & \gbsCell{detectionTint}{34}{53.02} & \gbsCell{detectionTint}{34}{55.52} & \gbsCell{detectionTint}{34}{57.98} & \gbsCell{detectionTint}{12}{60.06} & \gbsCell{detectionTint}{12}{61.82} \\
\noalign{\vskip .65pt}
\hspace{.6em}Resilient & \gbsCell{detectionTint}{34}{98.54} & \gbsCell{detectionTint}{78}{74.56} & \gbsCell{detectionTint}{78}{82.66} & \gbsCell{detectionTint}{78}{95.19} & \gbsCell{detectionTint}{78}{98.71} & \gbsCell{detectionTint}{78}{98.93} & \gbsCell{detectionTint}{78}{53.12} & \gbsCell{detectionTint}{78}{56.77} & \gbsCell{detectionTint}{78}{64.86} & \gbsCell{detectionTint}{78}{73.04} & \gbsCell{detectionTint}{78}{78.65} \\
\noalign{\vskip .65pt}
\hspace{.6em}Sparse-offset & \gbsCell{detectionTint}{78}{99.23} & \gbsCell{detectionTint}{12}{73.97} & \gbsCell{detectionTint}{12}{76.87} & \gbsCell{detectionTint}{34}{83.72} & \gbsCell{detectionTint}{34}{90.45} & \gbsCell{detectionTint}{34}{92.90} & \gbsCell{detectionTint}{12}{52.61} & \gbsCell{detectionTint}{12}{53.95} & \gbsCell{detectionTint}{12}{57.25} & \gbsCell{detectionTint}{34}{61.76} & \gbsCell{detectionTint}{34}{65.93} \\
\noalign{\vskip .65pt}
\addlinespace[3pt]
\multicolumn{12}{@{}l}{\bfseries\itshape Mean state estimation error} \\
\hspace{.6em}GBS & \gbsCell{errorTint}{78}{0.145} & \gbsCell{errorTint}{56}{0.205} & \gbsCell{errorTint}{100}{0.215} & \gbsCell{errorTint}{100}{0.178} & \gbsCell{errorTint}{100}{0.159} & \gbsCell{errorTint}{100}{0.154} & \gbsCell{errorTint}{34}{0.200} & \gbsCell{errorTint}{100}{0.232} & \gbsCell{errorTint}{100}{0.227} & \gbsCell{errorTint}{100}{0.218} & \gbsCell{errorTint}{100}{0.211} \\
\noalign{\vskip .65pt}
\hspace{.6em}$\chi^2$ / CUSUM & \gbsCell{errorTint}{56}{0.155} & \gbsCell{errorTint}{78}{0.200} & \gbsCell{errorTint}{56}{0.297} & \gbsCell{errorTint}{56}{0.413} & \gbsCell{errorTint}{34}{0.534} & \gbsCell{errorTint}{34}{0.658} & \gbsCell{errorTint}{100}{0.186} & \gbsCell{errorTint}{56}{0.258} & \gbsCell{errorTint}{56}{0.347} & \gbsCell{errorTint}{56}{0.443} & \gbsCell{errorTint}{56}{0.542} \\
\noalign{\vskip .65pt}
\hspace{.6em}Resilient & \gbsCell{errorTint}{34}{0.161} & \gbsCell{errorTint}{100}{0.199} & \gbsCell{errorTint}{78}{0.238} & \gbsCell{errorTint}{78}{0.199} & \gbsCell{errorTint}{78}{0.175} & \gbsCell{errorTint}{78}{0.173} & \gbsCell{errorTint}{56}{0.188} & \gbsCell{errorTint}{78}{0.238} & \gbsCell{errorTint}{78}{0.257} & \gbsCell{errorTint}{78}{0.247} & \gbsCell{errorTint}{78}{0.234} \\
\noalign{\vskip .65pt}
\hspace{.6em}Sparse-offset & \gbsCell{errorTint}{100}{0.140} & \gbsCell{errorTint}{56}{0.205} & \gbsCell{errorTint}{34}{0.328} & \gbsCell{errorTint}{34}{0.438} & \gbsCell{errorTint}{56}{0.500} & \gbsCell{errorTint}{56}{0.515} & \gbsCell{errorTint}{78}{0.187} & \gbsCell{errorTint}{34}{0.286} & \gbsCell{errorTint}{34}{0.389} & \gbsCell{errorTint}{34}{0.477} & \gbsCell{errorTint}{56}{0.542} \\
\noalign{\vskip .65pt}
\bottomrule
\end{tabular}

\end{table*}

{\color{revision}
This subsection compares GBS with representative detectors and estimators when neither the attacked sensor nor the attack type is known in advance. We consider the system
\[
 x_{t+1}=Ax_t+Bu_t+w_t,\qquad y_t=Cx_t+Du_t+v_t,
\]
with scalar state, two sensors, $A=0.8$, $B=0.2$, $C=(0.9,\,0.8)^\top$, $D=(0.1,\,0.2)^\top$, $x_0\sim\mathcal N(0,1)$, $Q=0.5$, $R=2I_2$, and horizon $N=20$. All methods use the known input $u_t=t$ for $t=0,\dots,N$.

In each trial, a randomly chosen sensor $j\in\{1,2\}$ is attacked by $$\tilde y_{t,j}=y_{t,j}+e_t ,$$ while the other reports nominal measurements. Neither $j$ nor the attack type is revealed to any method. The attack signal $e_t$ takes one of four forms: random interference, $e_t\sim\mathcal N(0,v)$ i.i.d.; constant bias, $e_t= h$ for $t\ge 10$; increasing bias, $e_t$ growing linearly from $0$ to $h$; and fixed-lag replay~\cite{mo2009replay}, $e_t=y_{t-T,j}-y_{t,j}$ for $t\ge 10$ with $T\in\{2,4,6,8,10\}$.}



The following detectors and estimators are considered for comparison.
\begin{itemize}
    \item The Gaussian--Bernoulli secure (GBS) estimator jointly estimates the observation indicators and the state trajectory.
    \item The $\chi^2$ detector compares the squared standardized Kalman innovation with $\alpha=6$.
    \item The CUSUM detector accumulates the standardized innovation in both directions and flags a measurement when either sum exceeds $6$.
    \item The resilient estimator tests the squared standardized innovation before each update and ignores observations that exceed $\alpha=6$.
    \item The sparse-offset estimator~\cite{farahmand2011} jointly fits the state trajectory and sparse offsets $o_{t,j}$ in standardized measurement units by adding $\lambda\sum_{t,j}|o_{t,j}|$ to the nominal cost. We use $\lambda=\sqrt{6}$ and flag measurements with nonzero estimated offsets.
\end{itemize}
{\color{revision}We evaluate each attack at five nonzero settings and use a common attack-free baseline, with 10,000 independent observation sequences per setting. The methods use the same state and noise realizations. Detection success is the fraction of sensor--time pairs correctly classified as attacked or attack-free, averaged over trials. Both missed attacks and false alarms count as errors. We measure estimation error by $\|\hat x_{0:N}-x_{0:N}\|_2/(N+1)$ and average it over trials.}

{\color{revision}Table~\ref{tab:sensor-performance} shows that the same GBS formulation handles diverse attack patterns without requiring prior knowledge of the attack type. Under weak attacks, the five methods perform similarly, and the estimation errors remain small even though the corrupted measurements are hard to distinguish from nominal ones. Under moderate and strong structured attacks, GBS generally achieves the best trade-off between detection and estimation accuracy. These results confirm the benefit of jointly estimating the state trajectory and the observation indicators.}

\subsection{Multiple Attacks in Multi-sensor System}
In multi-sensor systems, detecting attacks and ensuring resilient estimation becomes harder as more sensors are compromised. The difficulty arises from the fact that, as the volume of observations grows, malicious data can be more easily hidden among outliers—rare but significant deviations.

\begin{figure}[htbp]
    \centering
    \begin{subfigure}[b]{0.23\textwidth}
        \centering
        \includegraphics[width=\textwidth]{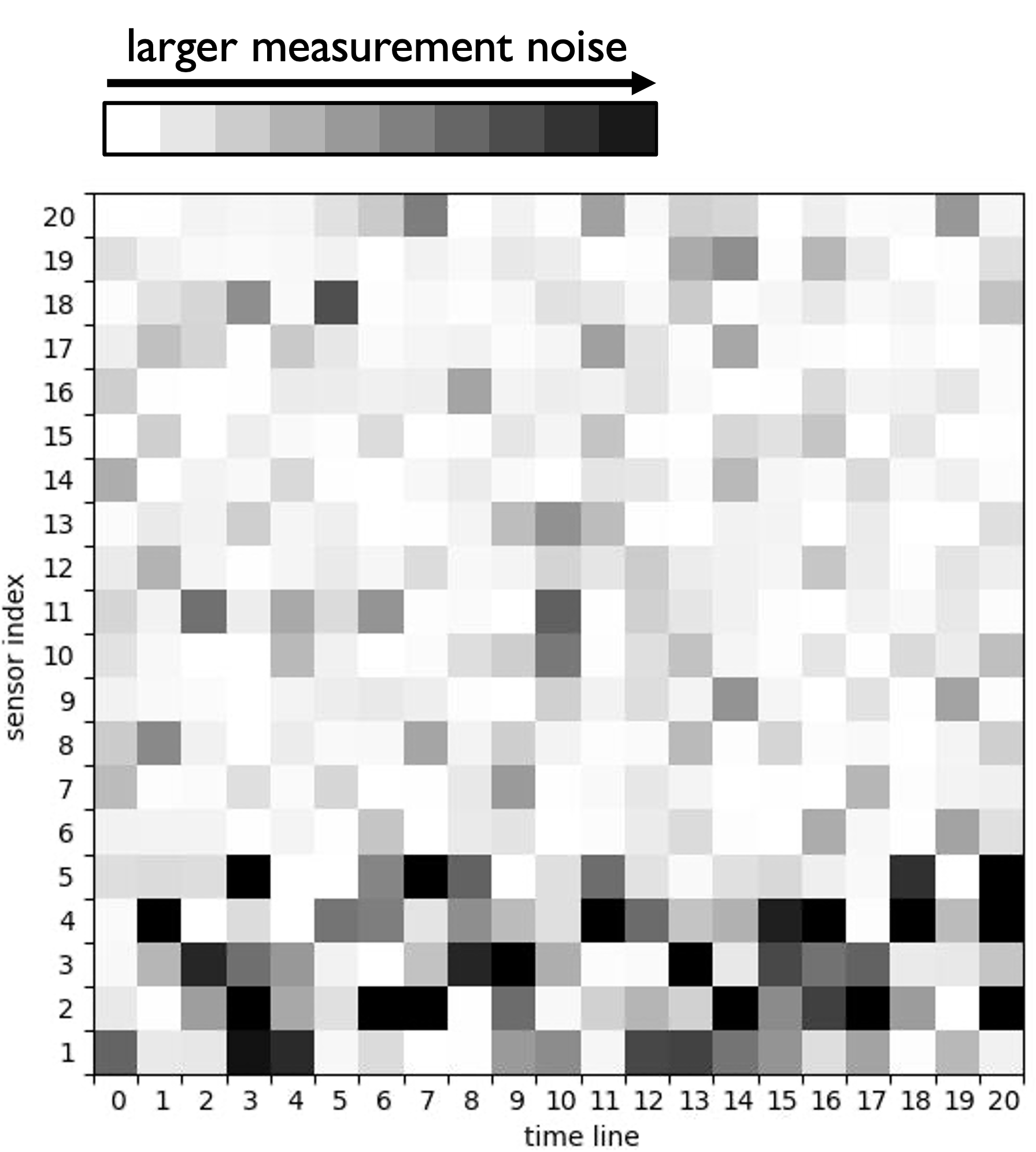}
        \subcaption{observation noise map}
        \label{multi_b}
    \end{subfigure}
    \begin{subfigure}[b]{0.23\textwidth}
        \centering
        \includegraphics[width=\textwidth]{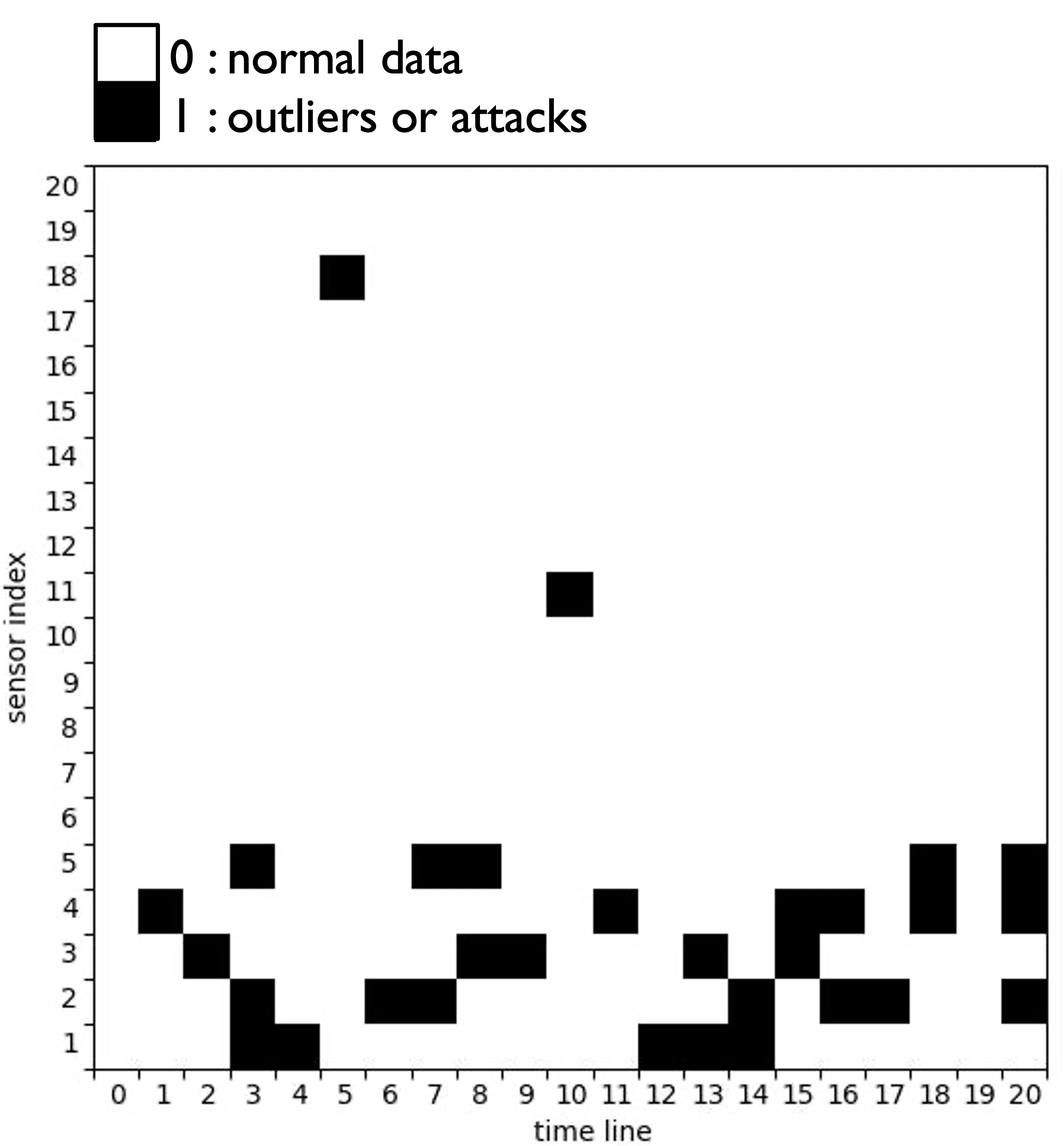}
        \subcaption{estimated indicators map}
        \label{multi_c}
    \end{subfigure}
    \caption{Characterization the measurement error by estimated observation indicators. (a) shows the true observation errors $ \|y_{i,j}-C_jx_i\|^2_{R_j^{-1}} $, which are continuous values. Sensor $1$ to $5$, under attack, generated several observations with significant errors. (b) shows our assessment of the reliability of each observation, which is boolean.}
\end{figure}

Consider a scalar system with $M=20$ sensors at interval $T=[0,20]$, where $x_0 \sim \mathcal{N}(0,1), A=1, Q=0.5.$
All sensors are supposed to satisfy following observation model. For each $i \in T$,
\begin{align*}
    y_{i,j} \sim \mathcal{N}(x_i, R_j = 20), \hspace{0.3cm}
        j=1,\cdots,20.
\end{align*}
Suppose we deploy noise interference on sensors indexed from $j^*=1$ to $5$. When these attacks last from $i=0$ to $20$, the actual observation model becomes:
\begin{align*}
    y_{i,j^*} \sim \mathcal{N}(x_i, \tilde{R}_{j^*} = 100), j^*=1,\cdots,5.
\end{align*}
Comparing Fig.~\ref{multi_b} and Fig.~\ref{multi_c}, their similarity illustrates that Gaussian-Bernoulli secure estimator effectively characterizes observation errors via indicators. {\color{revision}Counting the flagged readings identifies sensors with repeated deviations. The count alone does not determine their cause.}

\section{Conclusion}
We revisit the sensor attack problem from the perspective of system-environment interaction. Unlike conventional systems, interactive systems emphasize that observations not only contain information strongly correlated with the system’s internal state, but also carry information determined by the external environment. Under potential sensor attacks, {\color{revision}an observation depends on the system state during normal operation and can also contain an external disturbance when the sensor is compromised.} Therefore, in interactive systems, state estimation can be viewed as the process of decoding system information, while attack detection becomes the task of decoding environmental information.

There are two main challenges in decoding such observation information. The first challenge lies in how to effectively characterize the environmental attack information, especially when the attack model is unknown, to facilitate subsequent decoding. The second challenge is that information from these two distinct sources is often entangled in a single observation variable, making it necessary to decouple them before decoding.

The central motivation and contribution of our work is to address these two challenges. First, we propose the Bernoulli-Gaussian observation model assumption, where the system’s internal information is modeled using a Gaussian distribution, {\color{revision}and a Bernoulli indicator represents the occurrence of a deviation.} Based on this modeling assumption, we formulate a dual-variable optimization problem to jointly infer the system state and the observation indicators. Finally, we propose the GBS estimator, {\color{revision}which revises the observation indicators as it updates the state trajectory. The search terminates after finitely many changes. Its terminal estimate has a horizon-independent mean-square error bound if the retained clean observations satisfy the window observability condition.}

{Despite these contributions, several limitations remain and point to future directions. The current framework is established for linear systems with known parameters, and operates in a batch manner over a fixed horizon $[i, i+N]$, which may introduce detection delay in time-sensitive applications. Future work will extend the framework to nonlinear systems, develop an online sliding-window variant that re-optimizes the state and attack indicators over a moving horizon for real-time detection, design robust variants that accommodate model uncertainty and time-varying parameters, and extend the proposed method to robotic applications, particularly proprioception-based estimation of external contact forces.}

\appendices
\section{APPENDIX}
{\small
\subsection{Proof of Lemma~\ref{lemma:1}}\label{appendices:A}
{\color{revision}
Under the fixed-set nominal law the conditional density of $X$ is
proportional to $\exp[-\mathcal L_{\mathcal O}(X)/2]$. The prior and
transition residuals form an invertible block triangular transformation
of $X$, hence the quadratic has positive definite precision. Its unique
minimizer is the conditional mean. Gaussian conditioning on successive
retained observations followed by RTS smoothing evaluates that mean.

To obtain the stated linear map, let $K_t$ and $F_t$ denote the Kalman
and RTS gains for the prescribed schedule, with $K_t=0$ when
$t\notin\mathcal O$. Write $\bar x_t$ for the filtered mean and $e_t$
for the innovation relative to the prediction. Expanding
$\bar x_t=A\bar x_{t-1}+K_te_t$ from the zero prior mean yields
$\bar X=M_{\mathcal O}E$ and $Y=L_{\mathcal O}E$, where
\[
 (M_{\mathcal O})_{tj}=
 \begin{cases} A^{t-j}K_j,&j\le t,\\0,&j>t,\end{cases}
\]\[
 (L_{\mathcal O})_{tj}=
 \begin{cases} CA^{t-j}K_j,&j<t,\\I,&j=t,\\0,&j>t.\end{cases}
\]
Here $\bar X$ and $E$ stack the filtered states and innovations;
$L_{\mathcal O}$ is unit lower triangular. Write
$F_{t:j}=F_t\cdots F_{j-1}$ and $F_{t:t}=I$.
Expanding the backward recursion
$\hat x_{t|\mathcal O}=(I-F_t A)\bar x_t+F_t\hat x_{t+1|\mathcal O}$
from $\hat x_{N|\mathcal O}=\bar x_N$ gives
\[
 (H_{\mathcal O})_{tj}=
 \begin{cases}
 F_{t:j}(I-F_j A),&t\le j<N,\\
 F_{t:N},&j=N,\\
 0,&j<t.
 \end{cases}
\]
Combining these relations yields the formula in Lemma~\ref{lemma:1}.
An omitted observation has $K_j=0$ and therefore contributes to
neither the filtered nor the smoothed trajectory. \hfill$\square$
}

\subsection{Proof of Theorem~\ref{theorem:2}}\label{appendices:B}
{\color{revision}
Fix $\mathcal O$ and abbreviate $B=B_{\mathcal O}$, $F=F_{\mathcal O}$,
$b=b_{\mathcal O}$. Write $M=I+\eta H$ and
$U=M^{-1/2}(I-\eta F)M^{-1/2}$. The step-size condition gives
$0\preceq U\preceq M^{-1}\preceq qI$. If
$z_t=M^{1/2}(X^t-\hat X_{\mathcal O})$, subtracting the fixed-point
equation from \eqref{eq:pg-update} yields $z_{t+1}=U z_t$. Moreover
$\eta M^{-1/2}BM^{-1/2}=I-U$, so completing the square gives
\begin{equation*}
 E_t=\|X^t-\hat X_{\mathcal O}\|_B^2
     =\eta^{-1} z_t^T(I-U)z_t .
\end{equation*}
In an orthonormal eigenbasis of $U$, each term in $E_t$ is multiplied
by an eigenvalue squared at the next step. This proves $E_t\le q^{2t}E_0$.
For $v_t=BX^t-b$, the same identity gives
$E_t=v_t^T B^{-1}v_t\le\|v_t\|_2^2/\mu$, because
$B\succeq H\succeq\mu I$. \hfill$\square$
}

\subsection{Proof of Theorem~\ref{theorem:3}}\label{appendices:C}
{\color{revision}
Write $B=B_{\mathcal O}$, $D=\tilde C_i$, $r=r_i$, and
$S=R+DB^{-1}D^T$. For any trajectory increment $u$, expansion around
the minimizer of $\mathcal L_{\mathcal O}$ gives
\begin{align*}
 \mathcal L_{\mathcal O'}(\hat X_{\mathcal O}+u)
 -\mathcal L_{\mathcal O}(\hat X_{\mathcal O})
 &=\|u\|_B^2+\|r-Du\|_{R^{-1}}^2 .
\end{align*}
The unique minimizer is $u=B^{-1}D^TS^{-1}r$, as follows by substitution
in $(B+D^TR^{-1}D)u=D^TR^{-1}r$. At this point $r-Du=RS^{-1}r$,
and the minimum of the displayed expression is $r^TS^{-1}r$.
Evaluating it at $u=0$ and subtracting this minimum gives
\eqref{eq:warm-gap}. The iteration bound follows from \eqref{eq:inner-rate}.
\hfill$\square$
}

} 
{\color{revision}
\subsection{Proof of Theorem~\ref{thm:selected-stability}}
\label{app:selected-stability}
For a time set $\mathcal S$, define $\mathsf C_{\mathcal S}h=(C_Rh_j)_{j\in\mathcal S}$
and $H_{\mathcal S}=D^\top D+\mathsf C_{\mathcal S}^\top \mathsf C_{\mathcal S}$.
Let $e=\hat X-X$, $\tilde v_j=R^{-1/2}v_j$ and
$r_j=R^{-1/2}(y_j-C\hat x_j)$. The state optimality equations and the
identity $r_j=\tilde v_j-C_Re_j$ on $\mathcal G$ give
\begin{equation}
 H_{\mathcal G}e=-D^\top D X+\mathsf C_{\mathcal G}^\top\tilde v_{\mathcal G}
                      +\mathsf C_{\mathcal B}^\top r_{\mathcal B}.
 \label{eq:selected-normal}
\end{equation}
Block $j$ on the right has norm at most $z_j+b_j$. This calculation is
pathwise and therefore also applies when the data determine $\mathcal O$.

We first bound the spectrum of $H_{\mathcal G}$. On a fixed horizon,
$\mu=\lambda_{\min}(D^\top D)>0$ since $D$ is block triangular with
invertible diagonal blocks. An upper bound, uniform in $N$, is
\[
 M=\|P_0^{-1}\|_2+(1+\|A\|_2)^2\|Q^{-1}\|_2+c^2+1 .
\]
Under \eqref{eq:clean-window} a lower bound independent of $N$ follows
as well. For a window $I$ beginning at $s$, let $q_I$ stack $(Dh)_j$
for $s<j<s+L$, put $a=\max_{0\le k<L}\|A^k\|_2$ and
$u=La\|Q^{1/2}\|_2$. Expanding $h_j$ from $h_s$ gives
$h_j=A^{j-s}h_s+\sum_{i=s+1}^{j}A^{j-i}Q^{1/2}(Dh)_i$.
The stacked sum over $I$ has norm at most $u\|q_I\|_2$ by
Cauchy--Schwarz. Applying the window Gramian to $h_s$ therefore gives
\[
 \sqrt{\gamma}\|h_s\|_2
 \le \|\mathsf C_{\mathcal G\cap I}h\|_2+cu\|q_I\|_2 .
\]
With $d=\sqrt L\,a/\sqrt{\gamma}$ and $f=u(1+dc)$, the expansion
implies $\|h_I\|_2\le d\|\mathsf C_{\mathcal G\cap I}h\|_2+f\|q_I\|_2$.
Cover the horizon with consecutive disjoint full windows and, if
necessary, one last full window ending at $N$. Each sample is counted
at most twice. Squaring and summing yields
$\|h\|_2^2\le 2(d^2+f^2)h^\top H_{\mathcal G}h$. Hence in this case
we may take $\mu=[2(d^2+f^2)]^{-1}$.

Finally, $I-H_{\mathcal G}/M$ is block tridiagonal with spectral norm
at most $\rho=1-\mu/M<1$. Its $k$th power has a zero $(t,j)$ block
when $k<|t-j|$. Summing the Neumann series for $H_{\mathcal G}^{-1}$
gives $\|(H_{\mathcal G}^{-1})_{tj}\|_2\le\rho^{|t-j|}/\mu$.
Combining this inequality with \eqref{eq:selected-normal} proves
\eqref{eq:selected-decay}. At termination, dropping any retained
observation whose squared residual exceeded $\alpha$ would decrease
the objective even with $\hat X$ fixed. Single-indicator optimality
thus gives $b_j\le c\sqrt\alpha$. Minkowski's inequality and
$\sum_{j=0}^N\rho^{|t-j|}\le(1+\rho)/(1-\rho)$ yield
\eqref{eq:selected-rms}. If the window condition holds only on an
event $E_N$, the same argument bounds
$(\mathbb E[\mathbf 1_{E_N}\|\hat x_t-x_t\|_2^2])^{1/2}$ by the same
right-hand side. This last bound makes no assumption on selection
outside $E_N$. \hfill$\square$
}

\bibliographystyle{unsrt}
\bibliography{ref}
\end{document}